\documentclass{mscs}

\usepackage{epsfig}
\usepackage{epic}
\usepackage{eepic}
\usepackage{pictex}

\usepackage{amssymb}

\newtheorem{theorem}{Theorem}[section]
\newtheorem{define}[theorem]{Definition}
\newenvironment{definition}{\begin{define} \rm}{\end{define}}
\newtheorem{exa}[theorem]{Example}
\newenvironment{example}{\begin{exa} \rm}{\end{exa}}
\newtheorem{proposition}[theorem]{Proposition}

\newtheorem{lemma}[theorem]{Lemma}
\newtheorem{corollary}[theorem]{Corollary}

\newtheorem{claim}{Claim}
\newtheorem{exe}{Exercise}

\def\smallromani{\renewcommand{\theenumi}{\roman{enumi}}
        \renewcommand{\labelenumi}{(\theenumi)}}

\newcommand{\qed}{\hspace*{\fill}\nolinebreak\mbox{$\quad\square$}}

\newcommand{\rarrow}{\rightarrow}
\newcommand{\rrarrow}{\longrightarrow}
\newcommand{\labtran}[1]{\stackrel{#1}{\rrarrow}}
\newcommand{\slabtran}[1]{\stackrel{#1}{\Longrightarrow}}
\newcommand{\indrule}[2]{\frac{\raisebox{1ex}{$#1$}}{\raisebox{-1.5ex}{$#2$}}} 
 
\newcommand{\la}{\langle}
\newcommand{\ra}{\rangle}
\newcommand{\os}{[\![}
\newcommand{\cs}{]\!]}

\long\def\comment#1{}

\renewcommand{\arraystretch}{1.5}

\begin{document}

\title[Synchronous vs. Asynchronous $\pi$-calculus]{Comparing 
the Expressive Power of the Synchronous and the Asynchronous 
$\pi$-calculi\thanks{Work supported by the NSF-POWRE grant EIA-0074909.}}
\author[Catuscia Palamidessi]{Catuscia Palamidessi\\
INRIA Futurs\\
LIX, \'Ecole Polytechnique, 91128 Palaiseau Cedex, France\\
{\tt catuscia@lix.polytechnique.fr} }

\date{}
\maketitle

\renewcommand{\thefootnote}{\arabic{footnote}}


\begin{abstract}
The Asynchronous $\pi$-calculus, proposed by Honda and Tokoro (1991) and, independently, 
by Boudol (1992), is a subset of the $\pi$-calculus \cite{Milner:92:IC}
which contains no explicit operators for choice and output-prefixing.
The communication mechanism of this calculus, however, is powerful enough
to simulate output-prefixing, as shown 
by Honda and Tokoro (1991) and
by Boudol (1992), and input-guarded
choice, as shown by Nestmann and Pierce (2000).
A natural question arises, then, whether or not it is as expressive as
the full $\pi$-calculus. We show that this is not the case. More precisely, 
we show that there does not exist any uniform, fully distributed translation 
from the $\pi$-calculus into the asynchronous $\pi$-calculus, up to any 
``reasonable'' 
notion of equivalence. This result is based on the incapability of the
asynchronous $\pi$-calculus to break certain symmetries possibly present 
in the initial communication graph. By similar arguments, we prove a 
separation result between the $\pi$-calculus and CCS, 
and between the $\pi$-calculus
and the $\pi$-calculus with internal mobility, a subset of the $\pi$-calculus
proposed by Sangiorgi where the output actions can only transmit private names.
\end{abstract}

\section{Introduction}\label{Intro}

Communication is one of the fundamental concepts
in concurrent and distributed computation, and can be of many kinds: 
synchronous, asyn\-chro\-nous, one-to-one, one-to-many, etc. 
In this paper we focus on the distinction between 
synchronous and asyn\-chro\-nous. 
Synchronous communication is usually understood as {\it simultaneous} 
exchange of information between the partners; 
a``real life'' example is the telephone\footnote{Of course, 
communication by telephone can be thought as simultaneous 
only when the transmission time is negligible wrt the ``clock'' 
of the partners, which is a reasonable assumption when the 
partners are humans.}. 
In contrast, in asynchronous communication 
the action of sending a message and the action of reading it
usually take place at different times. 
An example is email. 
The advantages and disadvantages of the two
methods are well known: the first is more costly, 
because it requires the partners to synchronize to 
establish the communication, but then, once established, 
it is more effective. 

\subsection{Motivations}
A question which arises naturally is whether 
these two mechanisms are equivalent; i.e., 
whether the one can implement the other. 
One direction seems simple, at least in principle:
asynchronous communication can be simulated by inserting 
between each pair of communicating agents a ``buffer''  process, 
see for instance \cite{Milner:89:BOOK} and 
\cite{He:1990:IFIP}\footnote{Depending on the particular kind of 
asynchronous communication this implementation can be more or less
complicated; for instance if the order in which messages are sent 
is to be maintained, then we need a FIFO buffer, 
whose definition as a process requires guarded 
nondeterminism. For unordered communication we just need a
bag, which can be defined by using input, output, 
and replication only.}.
The other direction, on the contrary, is not clear and 
researchers in the field seem to have radically different 
opinions about it.

The motivation for this work arises from the attempt 
of solving, or at least clarifying, this question. 
In the author's opinion, the crucial point is which other 
mechanisms are available in combination with 
synchronous communication:
If the processes can make choices together,  
based on the information that they exchange 
simultaneously, then synchronous communication is intuitively more powerful.
This intuition is supported by the example of two people who try 
to take a common decision by using email instead of the telephone: 
If they act always in the same way, i.e. they send 
at the same time identical mails
and react in the same way to what they read, then
an agreement may be never reached. 
 
The $\pi$-calculus \cite{Milner:92:IC} is a convenient framework 
to study this problem. 
In fact, the $\pi$-calculus is a synchronous paradigm which contains an
``asynchronous'' fragment \cite{Boudol:92:REPORT,Honda:91:ECOOP}. 
We can thus work in a uniform context. But, more important, 
the asynchronous $\pi$-calculus is one of the richest paradigm 
for asynchronous-communication  concurrency
introduced so far, hence a negative result
regarding this language is more significant.

\subsection{Background}
The asynchronous $\pi$-calculus differs from the 
$\pi$-cal\-cu\-lus for the lack of the choice and the output prefix
operators. The underlying model of interaction among processes, however, 
is the same as in the $\pi$-calculus (the communication rule 
is based on  handshaking, i.e. on the simultaneous 
execution of complementary actions). 
The reason why it is considered asynchronous is 
that, due to the lack of output prefix, an output action 
can only be written ``in parallel'' with other activities.
More precisely, in the $\pi$-cal\-cu\-lus we can write
$P = \bar{x}.P'$ to represent a process $P$ that performs an output 
on channel $x$, and continues as $P'$ afterward, and we can write 
$Q = x.Q'$ to represent a process $Q$ that performs an input on $x$ and 
continues as $Q'$ afterward. (In the $\pi$-calculi input and output 
actions have parameters (names), 
but here for simplicity we omit them.) 
Furthermore, by using the restriction operator
we can enforce the synchronization 
of $P$ and $Q$ on $x$, so that the processes can 
proceed only when the communication along $x$ takes place. 
 From the point of view of the sender, the attempt to 
perform the output provokes the suspension of $P$, 
and the execution of the matching input action (reception) 
resumes $P$ at its continuation point $P'$. 
In the asynchronous $\pi$-calculus, on the contrary, 
we can write a sender process only 
in the form  $P = \bar{x}\,|\,P'$, 
where $|$ is the parallel operator. 
Since it is performed in parallel, 
the output action $\bar{x}$ does not 
automatically suspend the sender, 
and there is no primitive notion of continuation point. 
One can think of   $\bar{x}\,|\,P'$ 
as a  process which performs an output on $x$ 
at some unspecified moment 
and of the handshaking between $\bar{x}$ and $x.Q'$
as the moment in which the message is received. 
The reception enables the continuation point 
$Q'$ in the receiver, but it does not cause (directly) 
any resumption of activity in the sender\footnote{Note 
that this kind of communication is unordered
and that $(\bar{x}\,|\,P') \,|\, {x}.Q' $ is equivalent 
(modulo silent actions) 
to $\nu y(\bar{y}.P'  \,|\, B \,|\, {x}.Q')$ where $B$ is 
the ``bag'' process $!\,y.\bar{x}$ and $y$ is a fresh name.}.

Of course, the effect of the output prefix can be simulated by 
implementing a {\it rendez-vous} mechanism, in which the receiver 
sends, upon reception of the message, an acknowledgment to the sender, 
and the sender waits until it receives such acknowledgment. 
This kind of technique was in fact used by Boudol (1992) to define 
an encoding of the output prefix in the asynchronous 
$\pi$-calculus. Although the technique may be not surprising, 
the appeal and the novelty of this encoding consists in an
elegant use of the primitives for link mobility, 
thanks to which the translation can be defined in a compact and 
fully compositional way. Independently, 
Honda and Tokoro (1991) proposed an encoding even more 
compact, and also fully compositional, in which  
it is the  receiver which takes the 
initiative of synchronizing with the sender. 
It is probably fair to say that part of the success of the 
asynchronous $\pi$-calculus, at least in the early days, was due
to the encoding of Honda and Tokoro. 
Another important factor, 
of course, was that it could be implemented in a relatively simple 
and natural way. The first implementation of
the asynchronous $\pi$-calculus (actually a version of it, called PICT)
was developed by Pierce and Turner (1998). 
\nocite{PIERCE:98:MILNER}

Both the encodings of Boudol and of Honda and Tokoro work only
when the output prefix is not used in combination 
with the choice operator. In a subsequent paper 
Honda and Tokoro (1992) showed an encoding for the 
choice operator, but only for the simple case of 
the local (aka internal, or blind) choice. 
More recently, however, Nestmann and Pierce (2000)
\nocite{NestmannPierce:00:IC}
showed that input-guarded choice can also be encoded 
compositionally in the asynchronous $\pi$-calculus\footnote{Nestmann and 
Pierce actually were interested in obtaining a so-called 
{\it fully abstract} translation, and 
proposed two kinds of encoding. The first one is 
fully abstract wrt {\it weak bisimulation} (see, for 
instance, \cite{Milner:89:BOOK}), but it
introduces divergences. The second one is fully abstract only
wrt a weaker relation called 
{\it coupled simulation} \cite{Parrow:1992:CONCUR}, 
but it is divergence-free. 
Here we refer to the second encoding, 
as we take the point of view that a notion of encoding used to 
relate the expressive power of two languages should 
not introduce divergences.}.
This result was another fundamental 
contribution towards the affirmation of
the asynchronous $\pi$-calculus as a practically 
useful paradigm:  input-guarded choice 
is considered a very convenient mechanism for 
programming concurrent systems, as it allows 
a process to suspend on two (or more) 
alternative input channels, and to resume as soon 
as one of them  receives a datum.
A language without this feature will have to 
use busy waiting or risk that a process is 
stuck forever on the ``wrong'' channel. 
Most implementations of languages based on CSP 
\cite{Hoare:78:CACM,Hoare:85:CSP}, 
for instance, provide this construct as a primitive.        

After Nestmann and Pierce showed that the input-guarded choice does not 
represent a gap in the expressive power, several authors
have used a presentation of the asynchronous $\pi$-calculus 
which includes this construct as an operator of the language
(see for instance Boreale and Sangiorgi (1998a), and Amadio et al. (1998)).
\nocite{BS:98:TCS,Amadio:98:TCS}

With a slight modification, the translation of Nestmann and Pierce 
can be combined with the translation of Boudol, so to provide an 
encoding of both the output prefix and the input-guarded choice. 
Still, the $\pi$-calculus offers something more, namely the 
possibility for the two partners of the communication 
to make choices together. Consider for instance a process $P$ 
ready to send data (alternatively) 
to channels $x$, $y$ and $z$, and assume that 
process $Q$ is ready to receive data (alternatively) 
from $x$, $y$ and $w$. In the $\pi$-calculus, 
$P$ and $Q$ can be specified in such a way that 
they will choose $x$ or $y$ (arbitrarily), 
and neither of them will select the ``wrong channel'' 
(that is $z$ for $P$ and $w$ for $Q$). 
More precisely, what we need for this specification is 
the so-called {\it separate choice} construct: 
an output-guarded choice for $P$ 
($P = \bar{x}.P_1 + \bar{y}.P_2 + \bar{z}.P_3$) 
and an input-guarded choice for $Q$ ($Q = x.Q_1 + y.Q_2 + w.Q_3$).
Can such a construct be implemented in the asynchronous $\pi$-calculus? 
Clearly, one could implement it by backtracking from the wrong attempts, 
or by using a third process to coordinate the activities of $P$ and $Q$.
However, Nestmann (2000) \nocite{Nestmann:00:IC}
showed the surprising result that it is possible to encode such 
mechanism even without introducing divergences  
(such as those which would arise from
backtracking loops) and in a fully distributed way, 
i.e. without introducing
coordinator processes.

\subsection{The contribution of this paper}
At this point one may doubt that there be  any 
interesting cases of inter-process coordination, 
expressible in the $\pi$-calculus, that cannot be 
expressed in the asynchronous $\pi$-calculus as well. 
But in fact there are. Consider the following modification 
of previous example: 
A process $P'$ ready to output on $x$ and to input from 
$y$ (alternatively), and a process $Q'$
ready to output on $y$ and to input from $x$ (alternatively).
In the $\pi$-calculus we can simply use the so-called 
{\it mixed-choice} construct to define
$P' = \bar{x}.P_1 + y.P_2$ and $Q' = x.Q_1 + \bar{y}.Q_2$, 
and then enforce communication 
on $x$ and $y$. Apparently this example is similar 
to previous one, but there is a fundamental difference: 
$P'$ and $Q'$ are symmetric here, at least in their initial action, 
whereas in previous example $P$ initially
can only send and $Q$ can only 
receive. The encoding of Nestmann 
uses a protocol in which send and receive play 
completely different roles. 
As we will show in this paper, 
it is in general not possible to simulate the 
behavior of $P'$ and $Q'$ in 
the asynchronous $\pi$-calculus, 
and not even in the $\pi$-calculus 
with separate choice. 
Intuitively, the 
reason is that the agreement on the
 communication channel ($x$ or $y$) represents a 
situation in which the initial symmetry of 
$P'$ and $Q'$ is broken, and this cannot be achieved 
in a language which  supports separate choice only. 
For proving this result, we use techniques from 
the field of Distributed Computing. 
In particular we show that 
in certain symmetric networks 
it is not possible, with the 
separate-choice $\pi$-calculus, to solve the 
leader election problem, i.e. 
to guarantee that all processes will reach a common agreement 
(elect the leader) in a finite amount of time. 
It is possible, on the contrary, to solve this problem with the 
mixed-choice $\pi$-calculus. 

The use of this technique has been inspired by the work of 
Boug\'e (1988), \nocite{Bouge:1988:AI}
who showed a similar separation result 
concerning the $CSP$ \cite{Hoare:78:CACM,Hoare:85:CSP} and the fragment of 
$CSP$ with input-guards only, $CSP_{\it in}$.
However the mixed-choice $\pi$-calculus is a much richer language 
than $CSP_{\it in}$, and our result could not be derived  
from the result of Boug\'e. 
Interestingly, Boug\'e (1988) proved also that 
$CSP_{\it in}$ cannot be encoded 
into its choice-free fragment, by using similar techniques. 
This result does not hold
in the context of the $\pi$-calculus, as shown 
by  the above mentioned result of Nestmann and Pierce.
In the last section of this paper we will go into further details about 
the relation between \cite{Bouge:1988:AI} and our work.
  
Another question that we investigate in this paper is
to what extent the  $\pi$-calculus
is more powerful than its ``ancestor'' CCS \cite{Milner:89:BOOK}. 
For the sake of homogeneity, we consider the  value-passing version of CCS, 
in which the input and the output actions carry value parameters (messages).
This language, that we will call here CCS{$_{vp}$}, can be seen as 
a subset of the $\pi$-calculus (except for the relabeling 
operator, see Section~\ref{pi-hierarchy}).
The main difference is that in the $\pi$-calculus the messages are names 
which can later be used as communication channels, 
thus allowing to change dynamically the 
structure of the communication graph (link mobility). 
In combination with the mixed choice, link mobility is a 
very powerful feature 
for coordination of distributed activities, since it 
allows two remote processes, 
originally not directly connected, to establish a 
direct communication link $x$ 
and to take decision together 
on the basis of the synchronous exchange of information along $x$. 
By using a technique similar to the above one
(existence/non-existence of a certain symmetric electoral system)
we show that this capability makes the mixed-choice $\pi$-cal\-cu\-lus 
strictly more expressive than CCS{$_{vp}$}. 

Finally, we consider the expressiveness of 
the language $\pi_I$, the $\pi$-calculus with internal mobility, 
proposed by Sangiorgi (1996). \nocite{Sangiorgi:96:TCS} 
This is a subset of the 
(mixed-choice) $\pi$-calculus in which the parameters 
of output actions can 
be private names only. 
Because of this restriction, $\pi_I$ enjoys 
pleasant properties such as symmetric rules for input and output, and a 
much simpler theory for bisimulation equivalence. 
Boreale (1998) 
\nocite{Boreale:98:TCS}
has shown that the asynchronous version of 
$\pi_I$ is essentially as expressive as 
the asynchronous $\pi$-calculus. His encoding is based on the following idea: 
the main use of sending a non-private name in the $\pi$-calculus is when 
a process $P$ send a link $x$ to  $Q$ via an intermediate process $R$. 
$P$ and  $Q$ can then communicate directly on $x$. 
In $\pi_I$, $R$ could not send $x$ to $Q$, because it can send only his
private names. However, the above  mechanism can be simulated by installing in 
$R$ a repeater for $x$
which receives messages from $P$ and send them to 
$Q$, and viceversa. 
Intuitively, however, this idea does not work 
in the presence of (mixed) choice, 
because the choice tests the possibility of communication 
(guard), and selects the 
corresponding branch, in one single atomic step. 
In the encoding of Boreale
the communication along $x$ is not atomic anymore, 
and therefore the atomicity of 
the guarded-choice mechanism cannot be trivially encoded, 
unless the target language 
supports a much richer notion of choice, such as a choice depending on the 
possibility of performing a sequence of actions instead than a single action. 
By using an argument similar to the one illustrated above for
CCS{$_{vp}$}, we  will show that in fact the mixed-choice $\pi$-calculus 
cannot be encoded in $\pi_I$.  

\subsection{The notion of encoding}
In the whole discussion above we have used the existence or non-existence 
of an encoding as the criterion to compare the expressive power of 
two languages. 
The various encodings presented in literature, however,
satisfy different 
structural and semantic requirements. 
What should be the properties of a good notion 
of encoding, to use as a basis for 
defining the concept of expressive power? 
We do not have a definitive answer. 
In the context of this paper, since we are 
interested in proving negative results 
(i.e. non-existence of an encoding) we
consider a {\it minimal} set of requirements. 
More specifically, we require an encoding $\os\cdot\cs$
to be
\begin{itemize}
\item {\it uniform}, i.e. 
\begin{itemize}
 \item homomorphic wrt the parallel operator, namely 
       $\os P \,|\,Q\cs = \os P \cs \,|\,\os Q\cs$, and 
       \item renaming preserving, in the sense that for any permutation of 
       names $\sigma$ in the domain of the source language there exists a
       permutation of 
       names $\theta$ in the domain of the target language such that
       $\os \sigma(P) \cs = \theta(\os P \cs)$\footnote{Note that in 
       \cite{Palamidessi:97:POPL} 
       we had a stronger condition, 
       namely $\os  \sigma(P) \cs = \sigma(\os P \cs)$.
       We realized however that the latter condition would be too strong, 
       for instance it would make problematic the introduction of new names in 
       the translation.}.
\end{itemize}
\item {\it semantically reasonable}, 
i.e. preserving the relevant observables and 
 the termination properties. 
\end{itemize}
These conditions will be defined more precisely in Section~\ref{uniform encoding}.

The requirement of homomorphism wrt the parallel operator
ensures that two parallel processes are translated into two parallel 
processes, i.e. no coordinator is added by the translation. 
Therefore we can interpret this requirement as the
condition that the degree of distribution of the processes be 
maintained by the translation. This condition makes the notion of 
encoding suitable
to compare expressiveness of languages for distributed systems,
where processes are expected to coordinate without the help of a 
centralized control.

The requirement of renaming preserving
ensures that the translation does not depend on channel names. 
This condition seems natural if we want the encoding to preserve the 
portability of processes across the nodes of a distributed network.

All the  encodings discussed above satisfy the two criteria 
of uniformity and reasonableness, 
with the possible exception of the encoding of Boreale, for which the 
preservation of a reasonable semantics is an open question.
Boreale in fact has shown the correctness of his encoding 
only wrt barbed bisimulation, which is not sensitive to 
internal loops. 

\subsection{The $\pi$-calculus hierarchy}\label{pi-hierarchy}
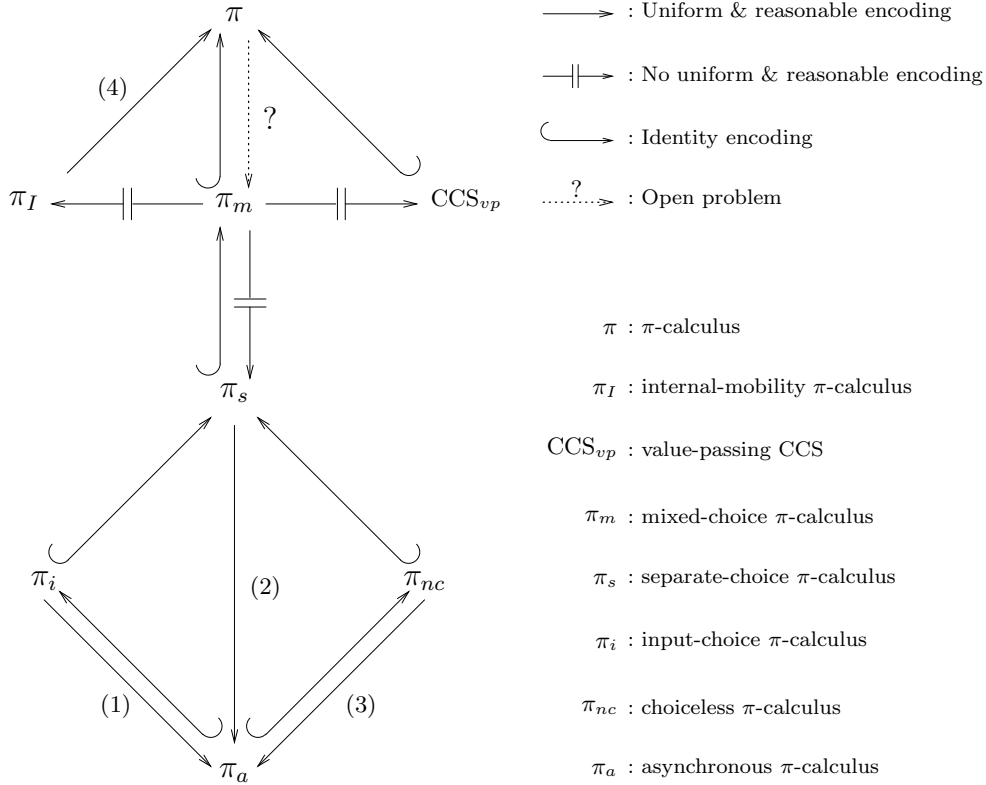
\begin{figure}
\begin{center}
\setlength{\unitlength}{0.00073333in}
\begingroup\makeatletter\ifx\SetFigFont\undefined%
\gdef\SetFigFont#1#2#3#4#5{%
  \reset@font\fontsize{#1}{#2pt}%
  \fontfamily{#3}\fontseries{#4}\fontshape{#5}%
  \selectfont}%
\fi\endgroup%
{\renewcommand{\dashlinestretch}{30}
\begin{picture}(6664,5588)(0,-10)
\path(1808,4161)(2307,4161)
\path(2319,4048)(2319,4275)
\path(2375,4048)(2375,4275)
\path(2375,4161)(2884,4161)
\path(2793.390,4138.350)(2884.000,4161.000)(2793.390,4183.650)
\path(1356,4161)(856,4161)
\path(847,4275)(847,4048)
\path(788,4275)(788,4048)
\path(788,4161)(279,4161)
\path(369.610,4183.650)(279.000,4161.000)(369.610,4138.350)
\path(1480,4355)(1480,5374)
\path(1502.650,5283.390)(1480.000,5374.000)(1457.350,5283.390)
\path(1583,2577)(1583,310)
\path(1560.350,400.610)(1583.000,310.000)(1605.650,400.610)
\path(1808,366)(2828,1387)
\path(2779.984,1306.890)(2828.000,1387.000)(2747.937,1338.906)
\path(1356,366)(335,1387)
\path(415.087,1338.945)(335.000,1387.000)(383.055,1306.913)
\path(2769,1614)(1751,2632)
\path(1831.087,2583.945)(1751.000,2632.000)(1799.055,2551.913)
\path(393,1614)(1412,2632)
\path(1363.906,2551.937)(1412.000,2632.000)(1331.890,2583.984)
\path(223,1329)(1412,139)
\path(1331.933,187.089)(1412.000,139.000)(1363.979,219.107)
\path(1480,3018)(1480,3990)
\path(1502.650,3900.390)(1480.000,3990.000)(1457.350,3900.390)
\path(1694,3970)(1694,3492)
\path(1583,3481)(1808,3481)
\path(1583,3424)(1808,3424)
\path(1694,3424)(1694,2916)
\path(1671.350,3006.610)(1694.000,2916.000)(1716.650,3006.610)
\dottedline{45}(1684,5327)(1684,4284)
\path(1661.350,4374.610)(1684.000,4284.000)(1706.650,4374.610)
\path(388,4382)(1407,5401)
\path(1363.945,5325.913)(1412.000,5406.000)(1331.913,5357.945)
\path(2775,4382)(1757,5401)
\path(1837.063,5352.906)(1757.000,5401.000)(1805.016,5320.890)
\path(2941,1329)(1751,139)
\path(1799.055,219.087)(1751.000,139.000)(1831.087,187.055)
\put(1751.000,423.000){\arc{161.220}{0.7854}{3.9270}}
\put(1412.000,423.000){\arc{159.812}{5.4889}{8.6305}}
\put(336.000,1670.000){\arc{159.812}{0.7765}{3.9181}}
\put(2826.500,1670.000){\arc{160.527}{5.5110}{8.6526}}
\put(1393.849,3020.613){\arc{172.547}{0.0535}{2.9596}}
 \put(2828.500,4443.000){\arc{160.515}{5.4934}{8.6350}}
\put(1393.664,4358.920){\arc{173.079}{0.0685}{2.9330}}
\put(600,4900){\makebox(0,0)[lb]{\small (4)}}
\put(1583,27){\makebox(0,0)[b]{\large $\pi_a$}}
\put(223,1400){\makebox(0,0)[b]{\large $\pi_i$}}
\put(2941,1400){\makebox(0,0)[b]{\large $\pi_{nc}$}}
\put(1583,5465){\makebox(0,0)[b]{\large $\pi$}}
\put(1583,4105){\makebox(0,0)[b]{\large $\pi_m$}}
\put(1583,2745){\makebox(0,0)[b]{\large $\pi_s$}}
\put(2987,4105){\makebox(0,0)[lb]{CCS{$_{vp}$}}}
\put(188,4116){\makebox(0,0)[rb]{\large $\pi_I$}}
\put(1785,4694){\makebox(0,0)[lb]{\large ?}}
\put(2375,480){\makebox(0,0)[lb]{\small (3)}}
\put(1694,1329){\makebox(0,0)[lb]{\small (2)}}
\put(847,480){\makebox(0,0)[rb]{\small (1)}}
\path(3790,5521)(4283,5521)
\path(4192.390,5498.350)(4283.000,5521.000)(4192.390,5543.650)
\path(3790,5075)(4000,5075)
\path(4000,4993)(4000,5158)
\path(4040,4993)(4040,5158)
\path(4040,5075)(4283,5075)
\path(4216.550,5058.385)(4283.000,5075.000)(4216.550,5091.615)
\path(3840,4614)(4283,4614)
\put(3837.280,4681.484){\arc{127.051}{1.6067}{4.5339}}
\path(4216.550,4597.010)(4283.000,4614.000)(4216.550,4630.990)
\dottedline{45}(3790,4173)(4271,4173)(4283,4173)
\put(3988,4208){\makebox(0,0)[lb]{?}}
\path(4192.390,4150.350)(4283.000,4173.000)(4192.390,4195.650)
%
\put(4394,5465){\makebox(0,0)[lb]{\footnotesize : Uniform \& reasonable encoding}}
\put(4394,5011){\makebox(0,0)[lb]{\footnotesize : No uniform \& reasonable encoding}}
\put(4394,4558){\makebox(0,0)[lb]{\footnotesize : Identity encoding}}
\put(4394,4128){\makebox(0,0)[lb]{\footnotesize : Open problem}}
\put(4322,3233){\makebox(0,0)[rb]{$\pi$}}
\put(4322,2779){\makebox(0,0)[rb]{$\pi_I$}}
\put(4322,2327){\makebox(0,0)[rb]{CCS$_{vp}$}}
\put(4322,1873){\makebox(0,0)[rb]{$\pi_m$}}
\put(4322,1420){\makebox(0,0)[rb]{$\pi_s$}}
\put(4322,966){\makebox(0,0)[rb]{$\pi_i$}}
\put(4322,513){\makebox(0,0)[rb]{$\pi_{nc}$}}
\put(4322,63){\makebox(0,0)[rb]{$\pi_a$}}
\put(4394,3233){\makebox(0,0)[lb]{\footnotesize : $\pi$-calculus}}
\put(4394,2779){\makebox(0,0)[lb]{\footnotesize : internal-mobility $\pi$-calculus}}
\put(4394,2327){\makebox(0,0)[lb]{\footnotesize: value-passing CCS}}
\put(4394,1873){\makebox(0,0)[lb]{\footnotesize : mixed-choice $\pi$-calculus}}
\put(4394,1420){\makebox(0,0)[lb]{\footnotesize : separate-choice $\pi$-calculus}}
\put(4394,966){\makebox(0,0)[lb]{\footnotesize : input-choice $\pi$-calculus}}
\put(4394,513){\makebox(0,0)[lb]{\footnotesize : choiceless $\pi$-calculus}}
\put(4394,63){\makebox(0,0)[lb]{\footnotesize : asynchronous $\pi$-calculus}}
\end{picture}
}
\caption{The $\pi$-calculus hierarchy. 
}\label{hierarchy}
\end{center}
\end{figure}

Figure~\ref{hierarchy} summarizes the results 
discussed above and introduces some terminology:
\begin{itemize}

\item
$\pi$ is the full $\pi$-calculus, as proposed in  \cite{Milner:92:IC}. 
In the $\pi$-calculus the choice operator ($+$) is {\it free}, in the sense 
that we can apply it to two arbitrary processes. For instance, we can 
write $(P|Q) + R$. 

\item $\pi_m$ stands for mixed-choice $\pi$, 
the subset of $\pi$ where the $+$ can occur
only among prefixed processes (e.g. $\Sigma_i \alpha_i.P_i$): 
the so called guarded choice operator. 
Here we say ``mixed choice'' to emphasize the fact that we can have 
both input and output (and silent) prefixes in the same guarded choice.
We also omit from $\pi_m$ the match and the mismatch operators. 
This is only because
the languages in the lower part of the diagram are traditionally 
presented without 
these operators, and  the result of non encoding between 
$\pi_m$ and $\pi_s$ would be 
weaker if $\pi_m$ had them. 
In general, the results obtained in this paper are independent 
from their presence or absence.  

\item $\pi_I$  is the internal-mobility $\pi$-calculus
introduced by Sangiorgi (1996): \nocite{Sangiorgi:96:TCS} in this language, 
an output parameter can only be written in the context of a restriction operator, 
e.g. $\nu x \, \bar{y} x. P$, also denoted as  $\bar{y}(x). P$. 
Another characteristics is that $\pi_I$ uses recursion instead than 
iteration. This is not accidental: in the context of $\pi_I$ 
iteration is strictly less expressive then recursion\nocite{Sangiorgi:96:TCS}. 
In $\pi$, on the contrary, recursion can be encoded by iteration \cite{Milner:93:BOOK}. 
For the rest $\pi_I$ is a subset of $\pi$, hence Milner's encoding of recursion 
extends naturally to an encoding of $\pi_I$ into $\pi$.

\item CCS{$_{vp}$} represents value-passing CCS 
without the relabeling operator.  
Value-passing means that actions have parameters, but unlike 
$\pi$ these parameters cannot be used as channels in other actions. 
If we consider values simply as symbols (i.e. we do not consider operators 
on values as part of the language), then value-passing CCS is a subset of $\pi$
except for the relabeling operator: such operator  
which does not exist in $\pi$ and according to Pugliese (1997) 
\nocite{Pugliese:97:PERS} it cannot be encoded either. 
For this reason we do not consider the relabeling operator here. 
It is worth noting, however, that the non-encoding of $\pi_m$ into 
CCS{$_{vp}$} does not depend on the absence of the the relabeling operator
(See Section~\ref{pi_m versus pi_I and CCS}). 

\item $\pi_s$ stands for separate-choice $\pi$, the subset of 
$\pi_m$ where the prefixes in a choice must 
be of the same kind
plus, possibly, $\tau$. Namely,  
in a guarded choice $\Sigma_i \alpha_i.P_i$  the $\alpha_i$'s 
which are not $\tau$ must all be either input or output action. 

\item $\pi_i$ represents input-choice $\pi$,
the subset of $\pi_s$ where there is no 
output prefix and only 
input actions can be used in a choice. Namely, in $\Sigma_i \alpha_i.P_i$
all the $\alpha_i$'s must be input actions. 

\item $\pi_{nc}$ stands for choiceless $\pi$, the subset of 
$\pi_s$ without choice, but with the
output prefix.
 
\item $\pi_a$ is the asynchronous $\pi$-calculus, 
namely the subset of $\pi_{nc}$ without 
output prefix. 
\end{itemize}

In Figure~\ref{hierarchy} some encodings are the obvious identity 
encodings holding between a language 
and a superset of it. As for the non-trivial encodings, 
(1) has been proposed by Nestmann and Pierce (2000), 
(2) has been proposed by Nestmann (2000), (3) 
represents the two encodings 
proposed by Honda and Tokoro (1991) and by Boudol (1992), 
and (4) is based on Milner's encoding of recursion 
into iteration. 
The three non-encodings are 
presented in this paper. 
The encoding of $\pi$ into $\pi_m$ is an open problem. It is 
likely that the free choice construct and the match and 
mismatch operators add 
expressive power. However, this result may 
not be obtainable with the weak requirements
that we have in this paper for the notion of encoding. 

Figure~\ref{hierarchy} shows only a few of the 
many variants of the $\pi$-calculus and of the many 
encodings and separation results which have been investigated 
in literature. We have considered here only the ones which to our opinion 
are the most relevant to the issues investigated in this paper. 
For a much more exhaustive overview of the variants of $\pi$ and of 
their expressive power we recommend the excellent book of 
Sangiorgi and Walker (2001). \nocite{Sangiorgi:01:BOOK}

\subsection{Organization of the paper}
The rest of the paper is organized as follows:  
next section recalls basic definitions. 
Section 3 reformulates in the setting of 
the $\pi$-calculus the notions of symmetric and electoral system. 
Section 4 shows the main result of the paper, i.e. 
the non-existence of symmetric electoral systems in 
the asynchronous $\pi$-calculus. Section 5 discusses 
existence of symmetric electoral systems for
the synchronous case, i.e. the $\pi$-calculus,
CCS{$_{vp}$}, and $\pi_I$.
Section 6 interprets previous results as 
non-encodability results. 
Section 7 discusses related work and concludes. 

\subsection{Relation with the previous version of this work}
A preliminary  version of this paper appeared in \cite{Palamidessi:97:POPL}. 
The principal differences in the present version  are: 
(a) The main separation result was 
previously shown between the mixed-choice $\pi$-calculus 
and the asynchronous $\pi$-calculus. 
Here we show that the separation lies exactly between the mixed-choice and the 
separate-choice (this could also be proved indirectly by combining 
the result in \cite{Palamidessi:97:POPL} and the result in 
\cite{Nestmann:00:IC}).
(b) We give the proof of the separation between the $\pi$-calculus and CCS$_{vp}$.
(In \cite{Palamidessi:97:POPL} this proof was only sketched.)
(c) We show an additional separation result, between the $\pi$-calculus and $\pi_I$, similar 
    to the one between the $\pi$-calculus and CCS$_{vp}$. 
(d) We consider a weaker condition on the notion of encoding 
    (and hence we strengthen the separation results).

\section{Preliminaries}
In this section we give the definitions of $\pi_m$, 
of $\pi_s$, 
and of the notion of hypergraph, which will be used to 
represent the communication structure of a network of processes. 

\subsection{The mixed-choice $\pi$-calculus}
We present here $\pi_m$, the mixed-choice $\pi$-calculus. 
This is a subset of the $\pi$-calculus which does not have
the match and mismatch operators, 
and in which the free choice is restricted to be guarded choice.
Note that  several recent papers adopt a presentation of the
$\pi$-calculus that actually coincides with $\pi_m$, see for instance
 \cite{BS:98:TCS,Sangiorgi:96:TCS}. 

Let  ${\cal N}$ be  a countable set of {\it names}, 
$x, y,\ldots$. The  set of prefixes,   
$\alpha, \beta,\ldots$,
and the set of $\pi$-calculus processes,   
$P,Q,\ldots$, are defined by the following 
abstract syntax: 

\[
\begin{array}{rlcl}
{\it Prefixes}&\alpha&\mbox{::=}&x(y)\;\;|\;\; \bar{x}y \;\;|\;\; \tau \\
{\it Processes}&P&\mbox{::=}& \sum_i\alpha_i.P_i \;\;|\;\; 
    \nu x P \;\;|\;\; P|P \;\;|\;\; !P
\end{array}
\]

Prefixes represent the basic actions of processes: 
$x(y)$ is the {\it input} of the (formal) name $y$ 
from channel $x$; 
$\bar{x}y$ is the {\it output} of the name $y$ 
on channel $x$; 
$\tau$ stands for any silent (non-communication) action. 

The process $\sum_i\alpha_i.P_i$ represents guarded (global) choice   
and it is usually assumed to be finite. 
We will use the abbreviations 
${\bf 0}$ ({\it inaction}) to represent the empty sum, 
$\alpha.P$ ({\it prefix}) to represent sum on one element only, and 
$P+Q$ for the binary sum. 
The symbols $\nu x$, $|$, and $!$ are the {\it restriction}, 
the {\it parallel}, and the {\it replication} operator, respectively. 

To indicate the structure of a process expression
we will use the following conventions: 
$P_0\,|\,P_1\,|\,P_2\,|\ldots|\,P_{k-1}$ stands for  
$(\ldots ((P_0\,|\,P_1)\,|\,P_2)\,|\ldots|\,P_{k-1})$, 
i.e. the parallel operator is left associative, 
and $\alpha_1.P_1 \,|\, \alpha_2.P_2$ stands for $(\alpha_1.P_1) | (\alpha_2.P_2)$, 
i.e. the prefix operator has precedence over $|$. In all other cases of 
ambiguity we will 
use parentheses.

The operators $\nu x$ and  $y(x)$ are $x$-{\it binders}, 
i.e. in the processes  $\nu x P$ and $y(x).P$ the occurrences 
of $x$ in $P$ are considered {\it bound}, 
with the usual rules of scoping. 
The set of the {\it free names} of $P$, i.e. those names which do 
not occur in the scope of any binder, 
is denoted by  ${\it fn}(P)$. 
The {\it alpha-conversion} of bound names is defined as usual, 
and the renaming (or substitution) $P\{y/x\}$ is defined as the 
result of replacing all occurrences of $x$ in $P$ by $y$, possibly 
applying alpha-conversion to avoid capture. 

The operational semantics is specified via a transition system 
labeled by {\it actions} $\mu, \mu'\ldots$. 
These are given by the following grammar: 
\[
\begin{array}{rlcl}
{\it Actions} &\mu &\mbox{::=}& x y \;\;|\;\; \bar{x} y \;\;|\;\; \bar{x}(y)
   \;\;|\;\; \tau
\end{array}
\]
Action $x y$ corresponds to the input prefix $x(z)$, where the 
formal parameter 
$z$ is instantiated to the actual parameter $y$ 
(see Rule {\sc I-Sum} in Table~\ref{TS}). 
Action $\bar{x} y$ correspond to the output of a free name. 
The {\it bound output} $\bar{x}(y)$ is introduced to model 
{\it scope extrusion}, i.e. the result of sending to another process 
a private ($\nu$-bound) name. 
The bound names of an action $\mu$, ${\it bn}(\mu)$,
are defined as follows:
${\it bn}(\bar{x}(y))=\{y\}$; 
${\it bn}(x y)={\it bn}(\bar{x}y)={\it bn}(\tau)=\emptyset$.
Furthermore, we will indicate by $n(\mu)$ all the 
{\it names} which occur in $\mu$.

In literature there are two 
definitions for the transition system of the $\pi$-calculus
which induce the so-called
{\it early} and {\it late} bisimulation semantics respectively. 
Here we choose to present the first one because the 
early strong bisimulation semantics is coarser than the late one.
Therefore, since our notion of reasonable semantics 
is coarser than strong bisimulation, a separation result with the 
early transition system is
more significant. 

The rules for the early semantics 
are given in Table~\ref{TS}. 
We use a congruence $\equiv$ 
and Rule {\sc Cong}
to simplify the presentation. 
We define this congruence as follows: 
\begin{enumerate}
\smallromani
\item \  $P\equiv Q$ if $Q$ can be 
obtained from $P$ by alpha-conversion, notation $P\equiv_\alpha Q$,
\item \  $(\nu x P) \,|\, Q\equiv \nu x (P\,|\,Q)$ if $x\not \in {\it fv}(Q)$ (scope expansion). 
\end{enumerate}
Some presentation of the labeled transition system 
of the $\pi$-calculus use a coarser definition of 
$\equiv$ obtained by adding other structural axioms
like the commutativity of $|$ (see for instance 
\cite{Milner:93:TCS}). Other presentations, like \cite{Sangiorgi:96:TCS}, 
define $\equiv$ as alpha conversion only, and use a congruence 
rule of the form 
\[
\indrule{P'\equiv P\ \ \ \ P\labtran{\mu}Q}
  {P'\labtran{\mu}Q}
\]
The reasons why we choose the above intermediate definition of $\equiv$ 
is because it seems to be the most 
suitable to prove the main theorem in Section \ref{nonexistence}. 
Given the way the systems we consider are structured,
we do not need the symmetric axiom for scope expansion.

\begin{table} 
\begin{center}
\begin{tabular}{|@{\ \ \ }ll@{\ \ }|}  
\hline
\mbox{}&\mbox{}
\\
{\sc I-Sum}
  &$\sum_i\alpha_i.P_i \labtran{x y} P_j\{y/z\}$
    \ \ \ \ $\alpha_j=x(z)$
\\
&\mbox{}
\\
{\sc O/$\tau$-Sum}
  &$\sum_i\alpha_i.P_i \labtran{\alpha_j} P_j$
    \ \ \ \ $\alpha_j=\bar{x}y$ or $\alpha_j=\tau$
\\
&\mbox{}
\\
{\sc Open}
 &$\indrule{P\labtran{\bar{x} y}P'}{\nu y P\labtran{\bar{x}(y)} P'}$
  \ \ \ \ $x\neq y$
\\
&\mbox{}
\\
{\sc Res}
 &$\indrule{P\labtran{\mu}P'}{\nu y P\labtran{\mu}\nu y P'}$
  \ \ \ \ $y\not\in n(\mu)$
\\
&\mbox{}
\\
{\sc Par}
 &$\indrule{P\labtran{\mu}P'}{P | Q\labtran{\mu}  P' | Q}$
  \ \ \ \ ${\it bn}(\mu)\cap{\it fn}(Q) =\emptyset$
\\
&\mbox{}
\\
{\sc Com}
 &$\indrule{P\labtran{x y}P'\ \ \ \ Q\labtran{\bar{x} y}Q'}
 {P | Q\labtran{\tau} P' | Q'}$
\\
&\mbox{}
\\
{\sc Close}
 &$\indrule{P\labtran{x y}P'\ \ \ \ Q\labtran{\bar{x}(y)}Q'}
  {P | Q\labtran{\tau} \nu y(P' | Q')}$
  \ \ \ \ $y \not \in {\it fn}(P)$
\\
&\mbox{}
\\
{\sc Rep}
 &$\indrule{P\,|\,!P\labtran{\mu}P'}
  {!P \labtran{\mu} P' }$
\\
&\mbox{}
\\
{\sc Cong}
 &$\indrule{P'\equiv P\ \ \ \ P\labtran{\mu}Q \ \ \ \ Q\equiv Q'}
  {P'\labtran{\mu}Q'}$
\\
&\mbox{} 
\\
\hline
\end{tabular}
\caption{The early-instantiation transition system for $\pi_m$.
The symmetric versions of {\sc Par}, {\sc Com} and {\sc Close}
are omitted.
}
\label{TS}
\end{center}
\end{table}

\subsection{The separate-choice $\pi$-calculus}

The separate-choice $\pi$-calculus, $\pi_s$, is the subset of $\pi_m$ in which 
output and input prefixes cannot be present in 
the same guarded choice. This restriction can be specified 
by the following modification in the grammar: 
\[
\begin{array}{rlcl}
{\it Input Prefixes}&\alpha^I&\mbox{::=}&x(y)\;\;|\;\; \tau \\
{\it Output Prefixes}&\alpha^O&\mbox{::=}&\bar{x}y \;\;|\;\; \tau \\
{\it Processes}&P&\mbox{::=}& \sum_i\alpha^I_i.P_i \;\;|\;\; \sum_i\alpha^O_i.P_i \;\;|\;\;
    \nu x P \;\;|\;\; P|P \;\;|\;\; !P
\end{array}
\]
The operational semantics of $\pi_s$ is the same as that of $\pi_m$, 
and it is described by the rules of Table~\ref{TS}.

\subsection{Hypergraphs and automorphisms}
In this section we recall the 
definition of {\it hypergraph}, which generalizes the concept of graph 
essentially by allowing an edge to connect more than 
two nodes. 

A hypergraph is a tuple $H=\la N, X, t \ra$
where $N,X$ are finite sets whose elements are called 
{\it nodes} and {\it edges} (or {\it hyperedges}) respectively, 
and $t$ ({\it type}) is a function which assigns to each $x\in X$ 
a set of nodes, representing the nodes {\it connected} by 
$x$. We will also use the notation $x:n_1,\ldots,n_k$ 
to indicate $t(x)= \{n_1,\ldots,n_k\}$. 

The concept of graph automorphism 
extends naturally to hypergraphs:
Given a hypergraph $H=\la N, X, t \ra$, 
an {\it automorphism} on $H$ is a pair $\sigma=\la \sigma_N,\sigma_X\ra$ 
such that $\sigma_N:N\rarrow N$ and $\sigma_X:X\rarrow X$ are 
permutations which preserve the type of edges, 
namely for each $x\in X$, if $x: n_1,\ldots,n_k$, 
then $\sigma_X(x):\sigma_N(n_1),\ldots,\sigma_N(n_k).$

It is easy to see that the {\it composition} of automorphisms, 
defined componentwise  as $\sigma\circ\sigma'= \la 
\sigma_N\circ\sigma'_N,\sigma_X\circ\sigma'_X\ra$,
is still an automorphism. Its identity is 
the pair of identity functions on $N$ and $X$, i.e.
${\it id}=\la {\it id}_N,{\it id}_X\ra$.
It is easy to show that the set of automorphisms
on $H$ with the composition forms a group.

Given $H$ and $\sigma$ as above, the {\it orbit} 
of $n\in N$ generated by $\sigma$ is defined as the set of nodes 
in which the various iterations of $\sigma$ map $n$, 
namely: 
\[
{\it O}_\sigma(n) = \{ n, \sigma(n), \sigma^2(n),\ldots , \sigma^{h-1}(n)\}
\]
\noindent
where $\sigma^i$ represents the composition of $\sigma$ with itself 
$i$ times, and $h$ is the least such that $\sigma^h={\it id}$.
It is possible to show that the orbits generated by $\sigma$ constitute 
a partition of $N$.

We say that an automorphism $\sigma$ is {\it well-balanced} if 
all its orbits have the same cardinality. 

\begin{example}
Figure~\ref{hypergraphs} illustrates various hypergraphs. 
Hypergraphs 1 and 2 correspond to standard graphs, in the sense that 
each of their edges connects only two nodes. In both of them we can define 
well-balanced automorphisms with 
\begin{itemize}
\item one single orbit with six nodes, or 
\item two orbits with three nodes each, or
\item three orbits with two nodes each
\end{itemize}
Of course, also the identity is a well-balanced automorphism 
(as in any hypergraph)
and in this case
it would have six orbits of cardinality one. 
 
Hypergraph 3 has six nodes and three edges, 
each of which connecting three nodes. This 
hypergraph has two 
well balanced automorphisms (apart from the identity), each with two orbits 
of cardinality three. 

Finally, Hypergraph 4 has seven nodes and three edges, 
each of which connecting four nodes. This hypergraph does not have 
any well-balanced automorphism except for the identity, 
because the central node has three incident edges while every other 
node has at most two incident edges. 
\end{example}

\begin{center}
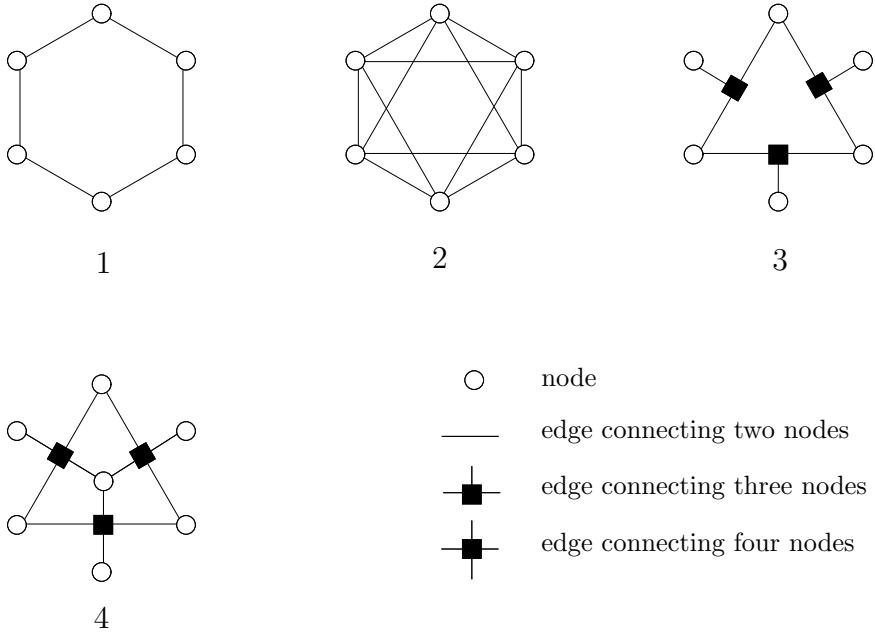
\begin{figure}
\setlength{\unitlength}{0.00065in}
\begingroup\makeatletter\ifx\SetFigFont\undefined%
\gdef\SetFigFont#1#2#3#4#5{%
  \reset@font\fontsize{#1}{#2pt}%
  \fontfamily{#3}\fontseries{#4}\fontshape{#5}%
  \selectfont}%
\fi\endgroup%
{\renewcommand{\dashlinestretch}{30}
\begin{picture}(6916,5000)(0,-10)
\path(3473,628)(3923,628)
\path(3473,628)(3923,628)
\path(3713,1288)(3713,1055)
\path(3713,1288)(3713,1055)
\path(3475,1528)(3925,1528)
\path(3475,1528)(3925,1528)
\path(3488,1078)(3938,1078)
\path(3488,1078)(3938,1078)
\path(6158,3402)(6158,3777)
\path(6833,4527)(6458,4302)
\path(91,1585)(775,1169)
\path(91,1585)(775,1169)
\path(1425,1585)(775,1169)
\path(1425,1585)(775,1169)
\path(775,502)(775,1152)
\path(775,502)(775,1152)
\path(765,1938)(123,828)(1406,828)(765,1938)
\blacken\path(6233,3702)(6233,3852)(6083,3852)
	(6083,3702)(6233,3702)
\path(6233,3702)(6233,3852)(6083,3852)
	(6083,3702)(6233,3702)
\blacken\path(6456,4228)(6586,4303)(6511,4433)
	(6381,4358)(6456,4228)
\path(6456,4228)(6586,4303)(6511,4433)
	(6381,4358)(6456,4228)
\put(4268,1513){\makebox(0,0)[lb]{\smash{{{\SetFigFont{10}{12.0}{\rmdefault}{\mddefault}{\updefault}edge connecting two nodes}}}}}
\put(4268,613){\makebox(0,0)[lb]{\smash{{{\SetFigFont{10}{12.0}{\rmdefault}{\mddefault}{\updefault}edge connecting four nodes}}}}}
\put(4268,1933){\makebox(0,0)[lb]{\smash{{{\SetFigFont{10}{12.0}{\rmdefault}{\mddefault}{\updefault}node}}}}}
\put(4268,1063){\makebox(0,0)[lb]{\smash{{{\SetFigFont{10}{12.0}{\rmdefault}{\mddefault}{\updefault}edge connecting three nodes}}}}}
\put(696,0){\makebox(0,0)[lb]{\smash{{{\SetFigFont{12}{14.4}{\rmdefault}{\mddefault}{\updefault}4}}}}}
\put(6113,2852){\makebox(0,0)[lb]{\smash{{{\SetFigFont{12}{14.4}{\rmdefault}{\mddefault}{\updefault}3}}}}}
\put(3396,2869){\makebox(0,0)[lb]{\smash{{{\SetFigFont{12}{14.4}{\rmdefault}{\mddefault}{\updefault}2}}}}}
\put(713,2835){\makebox(0,0)[lb]{\smash{{{\SetFigFont{12}{14.4}{\rmdefault}{\mddefault}{\updefault}1}}}}}
\path(3713,853)(3713,388)
\path(3713,853)(3713,388)
\blacken\path(3788,553)(3788,703)(3638,703)
	(3638,553)(3788,553)
\path(3788,553)(3788,703)(3638,703)
	(3638,553)(3788,553)
\path(3456,3409)(4098,4519)(2815,4519)(3456,3409)
\path(3465,4896)(2823,3786)(4106,3786)(3465,4896)
\blacken\path(3788,988)(3788,1138)(3638,1138)
	(3638,988)(3788,988)
\path(3788,988)(3788,1138)(3638,1138)
	(3638,988)(3788,988)
\path(6165,4896)(5523,3786)(6806,3786)(6165,4896)
\path(758,4902)(108,4527)(108,3777)
	(758,3402)(1408,3777)(1408,4527)(758,4902)
\blacken\path(1058,1272)(1188,1347)(1113,1477)
	(983,1402)(1058,1272)
\path(1058,1272)(1188,1347)(1113,1477)
	(983,1402)(1058,1272)
\path(5463,4509)(5838,4284)
\path(3458,4902)(2808,4527)(2808,3777)
	(3458,3402)(4108,3777)(4108,4527)(3458,4902)
\blacken\path(5912,4335)(5782,4410)(5707,4280)
	(5837,4205)(5912,4335)
\path(5912,4335)(5782,4410)(5707,4280)
	(5837,4205)(5912,4335)
\blacken\path(328,1347)(458,1272)(533,1402)
	(403,1477)(328,1347)
\path(328,1347)(458,1272)(533,1402)
	(403,1477)(328,1347)
\blacken\path(848,747)(848,897)(698,897)
	(698,747)(848,747)
\path(848,747)(848,897)(698,897)
	(698,747)(848,747)
\put(83,4527){\whiten\ellipse{150}{150}}
\put(83,4527){\ellipse{150}{150}}
\put(1433,4527){\whiten\ellipse{150}{150}}
\put(1433,4527){\ellipse{150}{150}}
\put(1433,3777){\whiten\ellipse{150}{150}}
\put(1433,3777){\ellipse{150}{150}}
\put(758,3402){\whiten\ellipse{150}{150}}
\put(758,3402){\ellipse{150}{150}}
\put(3458,4902){\whiten\ellipse{150}{150}}
\put(3458,4902){\ellipse{150}{150}}
\put(83,819){\whiten\ellipse{150}{150}}
\put(83,819){\ellipse{150}{150}}
\put(758,1944){\whiten\ellipse{150}{150}}
\put(758,1944){\ellipse{150}{150}}
\put(83,3777){\whiten\ellipse{150}{150}}
\put(83,3777){\ellipse{150}{150}}
\put(83,1569){\whiten\ellipse{150}{150}}
\put(83,1569){\ellipse{150}{150}}
\put(1433,1569){\whiten\ellipse{150}{150}}
\put(1433,1569){\ellipse{150}{150}}
\put(1433,819){\whiten\ellipse{150}{150}}
\put(1433,819){\ellipse{150}{150}}
\put(2783,3777){\whiten\ellipse{150}{150}}
\put(2783,3777){\ellipse{150}{150}}
\put(2783,4527){\whiten\ellipse{150}{150}}
\put(2783,4527){\ellipse{150}{150}}
\put(4133,4527){\whiten\ellipse{150}{150}}
\put(4133,4527){\ellipse{150}{150}}
\put(758,4902){\whiten\ellipse{150}{150}}
\put(758,4902){\ellipse{150}{150}}
\put(758,444){\whiten\ellipse{150}{150}}
\put(758,444){\ellipse{150}{150}}
\put(4133,3777){\whiten\ellipse{150}{150}}
\put(4133,3777){\ellipse{150}{150}}
\put(3458,3402){\whiten\ellipse{150}{150}}
\put(3458,3402){\ellipse{150}{150}}
\put(3732,1977){\whiten\ellipse{150}{150}}
\put(3732,1977){\ellipse{150}{150}}
\put(773,1167){\whiten\ellipse{150}{150}}
\put(773,1167){\ellipse{150}{150}}
\put(6158,4902){\whiten\ellipse{150}{150}}
\put(6158,4902){\ellipse{150}{150}}
\put(5483,3777){\whiten\ellipse{150}{150}}
\put(5483,3777){\ellipse{150}{150}}
\put(5483,4527){\whiten\ellipse{150}{150}}
\put(5483,4527){\ellipse{150}{150}}
\put(6833,4527){\whiten\ellipse{150}{150}}
\put(6833,4527){\ellipse{150}{150}}
\put(6833,3777){\whiten\ellipse{150}{150}}
\put(6833,3777){\ellipse{150}{150}}
\put(6158,3402){\whiten\ellipse{150}{150}}
\put(6158,3402){\ellipse{150}{150}}
\end{picture}
}
\caption{Examples of hypergraphs.}
\label{hypergraphs}
\end{figure}
\end{center}

\section{Electoral and Symmetric systems}

In this section we adapt to  the $\pi$-calculi
(a simplified version of)
the notions of electoral system and symmetric network as given 
by Boug\'e (1988). 

\subsection{Election of a leader in a network}
We first need to introduce  the concepts of 
{\it network}, 
{\it network computation}
and the 
{\it projection} of a computation over a component of the network.

A network represents a system of parallel processes with possibly some 
top-level applications of the restriction operators. The only difference 
between the notion of network and that of process is that in a network 
we want to represent explicitly the intended distribution. For instance, 
the process $((P_0\,|\,P_1)\,|\,P_2)$ (which, because of our associativity convention, 
we can write as $P_0\,|\,P_1\,|\,P_2$), may be interpreted as a network of three parallel 
processes, $P_0$, $P_1$ and $P_2$, or as a network of two parallel processes 
$P_0|P_1$ and $P_2$. 
Formally, a network is defined as a tuple of the form 
\begin{equation} \label{network}
 \la \la x_0, x_1, \ldots, x_{n-1}\ra , \la P_0, P_1, \ldots, P_{k-1}\ra \ra
\end{equation}
where the $P_i$'s are processes meant to run in parallel and the $x_i$'s are 
names which are meant to be globally bound (i.e. bound at the top 
level in the whole system). 
More precisely, the network (\ref{network}) is meant to represent the process
\begin{equation}\label{process}
P = \nu x_0\ \nu x_1 \ldots  \nu x_{n-1} (P_0\,|\,P_1\,|\ldots|\,P_{k-1})
\end{equation}
It will be convenient to assume that the 
bound names in $P_0, P_1, \ldots, P_{k-1}$ are 
different from each other, from  $x_0, x_1, \ldots, x_{n-1}$, and from 
all the free names (bound-names convention). 

 From now on we will use this process notation to represent the
network, with the convention that whenever we write 
an expression like that in (\ref{process})
we mean that the network is constituted
exactly by the $k$ processes $P_0,P_1,\ldots,P_{k-1}$. 
We will also use $[Q]$ to denote a process of the form 
$\nu x_0\ \nu x_1 \ldots  \nu x_{n-1} \ Q$. 
Thus whenever the specific bound names are not relevant 
we will simply represent the network above as 
\[
[P_0\,|\,P_1\,|\ldots|\,P_{k-1}].
\]

A computation $C$ for a  network is a  (possibly $\omega$-infinite)
sequence of transitions
\[
\begin{array}{lcl}
[P_0\,|\,P_1\,|\ldots|\,P_{k-1}]
 &\labtran{\mu^{0}}
   & [P^{1}_0 \,|\, P^{1}_1\,|\ldots|\,P^{1}_{k-1}]
   \\
 &\labtran{\mu^{1}} 
   & [P^{2}_0 \,|\, P^{2}_1\,|\ldots|\,P^{2}_{k-1}]
   \\
 &\vdots
   \\
 &\labtran{\mu^{n-1}} 
   & [P^{n}_0 \,|\, P^{n}_1\,|\ldots|\,P^{n}_{k-1}]
   \\
 & (\;\labtran{\mu^{n}}\;\;
   &\ldots\;) 
\end{array}
\]
\noindent
with $n\geq 0$.
Note that at each computation step we may need to apply the {\sc Cong}
rule on the right side of the transition in order:
\begin{itemize}
\item  to maintain 
the bound-names convention, i.e. to keep the bound names 
different from each other and from the
free names, and 
\item to maintain the parallel structure of the network.
In fact a transition generated by the {\sc Close} rule
would group a set of processes in the 
scope of a restriction operator, thus we need the {\sc Cong} 
rule with the axioms for scope expansion 
to bring the restriction operator at the top 
level and re-establish the number of components to $k$.
\end{itemize}

We will represent a computation like the above 
also by $C: P\slabtran{\tilde\mu}P^{n}$
(by $C: P\slabtran{\tilde\mu}$ if it is infinite),
$\tilde\mu$ being the sequence 
$\mu^{0}\mu^{1}\ldots\mu^{n-1} (\mu^{n}\ldots)$, and 
$P^n$ being the process $[P^{n}_1 \,|\, P^{n}_2\,|\ldots|\,P^{n}_k]$.
The relation $C\prec C'$ ($C'$ {\it extends} $C$) is defined as usual: 
let $C: P\slabtran{\tilde\mu}P^{n}$. Then $C\prec C'$
iff there exists $C'':P^{n}\slabtran{\tilde\mu'}P^{n+n'}$ with $n'\geq 1$,
or $C'':P^{n}\slabtran{\tilde\mu'}$, and
$C'= CC''$ (identifying the two occurrences of $P^n$). 
We will denote by $C'\setminus C$ the continuation $C''$.
Note that according to this definition 
infinite computations cannot be extended. This is consistent with the 
fact that we admit only $\omega$-infinite (i.e. not transfinite) 
computations.

Given $P$ and $C$ as above, the projection 
of $C$ over $P_i$, 
${\it Proj}(C,i)$,
is defined as the ``contribution'' of $P_i$ 
to the computation. 
More formally, ${\it Proj}(C,i)$
is the sequence of steps
\[
\begin{array}{c}
P_i\;\;\slabtran{\tilde\mu^{0}}\;\;Q_i\;\;\\
P^1_i\;\;\slabtran{\tilde\mu^{1}}\;\;Q^1_i\;\;\\
P^2_i\;\;\slabtran{\tilde\mu^{2}}\;\;Q^2_i\;\;\\
\vdots\;\;\\
P^{n-1}_i\;\;\slabtran{\tilde\mu^{n-1}}\;\;Q^{n-1}_i\\
\left( \begin{array}{c} 
       P^{n}_i\;\;\slabtran{\tilde\mu^{n}}\;\;Q^{n}_i\\
       \vdots
       \end{array}
\right)
\end{array}
\]
\noindent 
where the definition of $P^m_i\;\;\slabtran{\tilde\mu^{m}}\;\;Q^m_i$
depends on the proof tree $T$ which generates 
the transition step 
$[P^{m}_0 \,|\, P^{m}_1\,|\ldots|\,P^{m}_{k-1}]
 \labtran{\mu^{m}} 
[P^{m+1}_0 \,|\, P^{m+1}_1\,|\ldots|\,P^{m+1}_{k-1}]$:

\begin{itemize}
\item If $P^m_i$ is {\it active} during the transition, namely 
$T$ contains a node of the form 
$P^m_i\;\;\labtran{\mu}\;\;R$, then 
\[
(P^m_i\;\;\slabtran{\tilde\mu^{m}}\;\;Q^m_i) \ \ = 
\ \ (P^m_i\;\;\labtran{\mu}\;\;R)
\]
Note that either $P^m_i$ is the only process active during this transition, 
and in this case $\mu^{m}=\mu$, or $P^m_i$ is involved in a communication step 
(i.e. it is in the premise of a rule {\sc Com} or {\sc Close})
and in this case $\mu^{m}=\tau$ and $\mu$ is a communication action. 
Note also that $Q^m_i$ and $P^{m+1}_i$ may be different because 
$T$ may contain at some lower level an  application of the {\sc Cong} rule. 
However, $Q^m_i$ can only differ from $P^{m+1}_i$ 
for the presence of restriction operators and/or some renaming.

\item If $P^m_i$ is {\it idle} during the transition, namely 
$T$ does not contain any node of the form 
$P^m_i\, \labtran{\mu}\, R$, then 
$P^m_i\,\slabtran{\tilde\mu^{m}}\,Q^m_i$ is empty, 
namely $Q^m_i=P^{m}_i=P^{m+1}_i$ and $\tilde\mu^{m}$ is empty.
\end{itemize}

Note that the notation ${\it Proj}(C,i)$ is not 
accurate: the projection is not a 
function of $C$, but rather of the sequence 
of proof trees which generate $C$. 
However this distinction is inessential here.

In order to define the notion of electoral system
we assume the existence of a special channel 
{\it out} to be used for communicating with
the ``external world'', and therefore free (unbound). 
Furthermore we assume 
that the set of names $\cal N$ contains a special subset
equipped with a one-to-one mapping with the natural numbers, 
which we will use to identify the individual processes 
in a network. For the sake of simplicity we shall denote 
these names directly by natural numbers, but one should keep in mind that 
this is just a notation, i.e. we are not adding any arithmetical capability 
to the calculus. We will assume, without loss of generality, that 
these names are not used as bound names.

Intuitively an electoral system has the 
property that at each possible run
the processes will agree sooner or later on 
``which of them has to be the leader'', 
and will communicate this decision to the 
``external world'' by using the  special channel {\it out}. 

\begin{definition}\label{electoral}{\bf (Electoral system)}
A network $P=[P_0\,|\,P_1\,|\ldots|\,P_{k-1}]$
is an electoral system if 
for every computation $C$ for $P$ 
there exists an extension $C'$ of $C$ 
and there exists $n\in\{0,1,\ldots,k-1\}$
(the ``leader'')
such that for each $i\in \{0,1,\ldots,k-1\}$
the projection ${\it Proj}(C',i)$
contains one output action 
of the form $\overline{\it out}\,n$, and 
no extension of $C'$ contains
any other action of the form $\overline{\it out}\,m$, 
with $m\neq n$.
\end{definition}

Note that for such a system 
an infinite computation $C$ must contain already
all the output actions of each process
because $C$ cannot be extended.  

\subsection{Symmetric networks}

In order to define the notion of {\it symmetric network},
we have to consider its communication structure, 
which we will represent as an hypergraph.
Intuitively the nodes represent the 
processes, and the edges represent the communication channels
connecting the processes.
We exclude the special channel {\it out} since processes cannot 
use it to communicate with each other. 

\begin{definition}{\bf (Hypergraph associated to a network)}
Given a network 
$P= [P_0\,|\,P_1\,|\ldots|\,P_{k-1}]$, 
the hypergraph associated to $P$ 
is $H(P)=\la N,X,t\ra$ with $N=\{0,1,\ldots,k-1\}$, 
$X={\it fn}(P_0\,|\,P_1\,|\ldots|\,P_{k-1})\setminus\{{\it out}\}$, and 
for each $x\in X$, $t(x)= \{ n\,|\, x\in{\it fn}(P_n)\}$.
\end{definition}

We extend now the notion of automorphism to networks so to take into account
the use of the special names as process identifiers, and the role 
of the binders at the top level. 

\begin{definition}{\bf (Network automorphism)}
Given a network 
\[P= \nu x_0\ \nu x_1 \ldots  \nu x_{n-1} (P_0\,|\,P_1\,|\ldots|\,P_{k-1})\]
let $H(P)=\la N,X,t\ra$ be the hypergraph associated to $P$. 
An automorphism on $P$ is any automorphism $\sigma = \la \sigma_N, \sigma_X\ra$ 
on $H(P)$ which satisfies the following additional conditions: 
\begin{itemize}
\item $\sigma_X$ coincides with $\sigma_N$ on $N \cap X$, i.e. for every 
$n\in N\cap X$  we
have $\sigma_X(n) = \sigma_N(n)$ (remember that $N$ is a set of natural 
numbers and that the natural numbers are assumed to represent also a 
special subset of names).
\item $\sigma_X$ must preserve the distinction between free and bound names, 
i.e. 
\[
x\in \{x_0,x_1, \ldots , x_{n-1}\} \mbox{ \ if and only if \  }
\sigma(x)\in \{x_0,x_1, \ldots ,x_{n-1}\}, \mbox{ \ for each \  } x \in X
\]
\end{itemize}
\end{definition}

Thanks to the fact that $\sigma_N(\cdot)$ and $\sigma_X(\cdot)$ coincide 
on the intersection of their domains, 
we can simplify the notation and use $\sigma(\cdot)$ to represent both 
$\sigma_N(\cdot)$ and $\sigma_X(\cdot)$. We will also, with a slight abuse of notation, 
use the hypergraph $H$ to denote the domain of $\sigma_X(\cdot)$, i.e. the set 
$N\cup X$.
 
Intuitively, a network $P$ is symmetric with respect to an 
automorphism $\sigma$ iff  for each $i$
the process associated to the node 
$\sigma(i)$ is i\-den\-ti\-cal (modulo alpha-conversion) to the
process obtained by $\sigma$-renaming the process 
associated to the node $i$.  

The notion of $\sigma$-renaming is the obvious extension of 
the standard notion of renaming (see the preliminaries). 
More formally, given a process $Q$, first apply alpha-conversion 
so to rename all bound names into fresh ones, 
extend $\sigma$ to be the identity on these new names, and define 
$\sigma(Q)$ by structural induction as indicated below.
\[
\begin{array}{rcl}
\sigma(\tau)&=&\tau\\
\sigma(x(y))&=&\sigma(x)(y)\\
\sigma(\bar{x}y)&=&\overline{\sigma{\scriptstyle (}x{\scriptstyle )}}\sigma(y)\\
\sigma(\sum_i\alpha_i.P_i) &=& \sum_i\sigma(\alpha_i).\sigma(P_i)\\
\sigma(\nu x P) &=&\nu x\; \sigma(P)\\
\sigma(P \,|\, Q) &=&\sigma(P) \,|\, \sigma(Q)\\
\sigma(! P) &=&! \sigma(P)
\end{array}
\]
Furthermore we need to define the application of 
$\sigma$ on the actions. For $\tau$ and $\bar{x}y$ 
the definition is the same as above. For the other actions we have:
\[
\begin{array}{rcl}
\sigma(x y)&=&\sigma(x)\sigma(y)\\
\sigma(\bar{x}(y))&=& \overline{\sigma{\scriptstyle (}x{\scriptstyle )}}(y)
\end{array}
\]

We are now ready to give the formal definition of symmetric network:

\begin{definition}{\bf (Symmetric network)}
Consider a  network $P=[P_1\,|\,P_2\,|\ldots|\,P_k]$, 
let $H(P)=\la N, X,t\ra$ be its associated hypergraph, 
and let $\sigma$ be an automorphism on $P$. 
We say that $P$ is {\it symmetric wrt $\sigma$}
iff for each node $i\in N$, 
$P_{\sigma(i)}\equiv_\alpha\sigma(P_i)$ holds.
We also say that 
$P$ is {\it symmetric} if it is symmetric  wrt all the automorphisms 
on $H(P)$.
\end{definition}

Note that if $P$ is symmetric wrt $\sigma$ then 
$P$ is symmetric wrt all the powers of $\sigma$.

\section{Non existence of symmetric electoral 
systems in $\pi_s$}\label{nonexistence}

In this section we present our first result,
which says that
for certain communication graphs
it is not possible to 
write in $\pi_s$ a symmetric network 
solving the election problem. 

\begin{figure}
\begin{center}
\setlength{\unitlength}{3000sp}%
\begingroup\makeatletter\ifx\SetFigFont\undefined%
\gdef\SetFigFont#1#2#3#4#5{%
  \reset@font\fontsize{#1}{#2pt}%
  \fontfamily{#3}\fontseries{#4}\fontshape{#5}%
  \selectfont}%
\fi\endgroup%
\begin{picture}(2400,2460)(2401,-3586)
\thinlines
\put(3451,-1411){\vector(-1,-1){825}}
\put(3751,-1411){\vector( 1,-1){825}}
\multiput(2626,-2536)(86.84211,-86.84211){10}{\line( 1,-1){ 43.421}}
\put(3451,-3361){\vector( 1,-1){0}}
\multiput(4576,-2536)(-86.84211,-86.84211){10}{\line(-1,-1){ 43.421}}
\put(3751,-3361){\vector(-1,-1){0}}
\put(3601,-1261){\makebox(0,0)[b]{$P$}}
\put(2401,-2461){\makebox(0,0)[lb]{$Q$}}
\put(4801,-2461){\makebox(0,0)[rb]{$R$}}
\put(4201,-1711){\makebox(0,0)[lb]{$z w$}}
\put(3001,-1711){\makebox(0,0)[rb]{$\bar{x}[y]$}}
\put(4276,-3136){\makebox(0,0)[lb]{$\bar{x}[y]$}}
\put(2926,-3136){\makebox(0,0)[rb]{$z w$}}
\put(3601,-3586){\makebox(0,0)[b]{$S$}}
\end{picture}
\caption{An illustration of Lemma~\ref{confluence}. }\label{confluence_figure}
\end{center}
\end{figure}
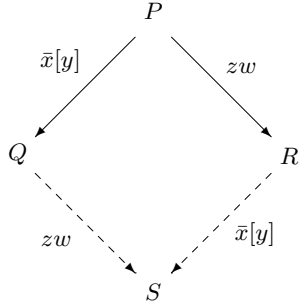

We first need to show that the $\pi_s$ 
enjoys a certain kind of {\it confluence} property: 

\begin{lemma}\label{confluence}
Let $P$ be a process in $\pi_s$. Assume that 
$P$ can make two transitions 
$P\labtran{\bar{x}[y]}Q$ and $P\labtran{z w}R$,
where $\bar{x}[y]$ stands for an output action either bound ($\bar{x}(y)$) or unbound ($\bar{x} y$).
Then there exists 
$S$ such that $Q\labtran{z w}S$ and 
$R\labtran{\bar{x}[y]}S$ (see Figure~\ref{confluence_figure}).
\end{lemma}

\noindent
{\bf Proof}
Observe that $x$ and $z$ must be free names in $P$.
The rule which has produced the $\bar{x}[y]$ transition can be only {\sc O/$\tau$-Sum},
{\sc Open}, {\sc Res}, {\sc Par}, {\sc Rep}, or {\sc Cong}. 
In the last five cases the 
assumption is again a $\bar{x}[y]$ transition. 
By repeating this reasoning (descending the tree), 
we must arrive to a leaf transition of the form
\[
P^O=\sum_i\alpha^O_i.P_i \labtran{\alpha^O_j} P_j \;\;\mbox{ where } \alpha^O_j = \bar{x} y
\] 
Analogously, the rule which has produced the $z w$ transition
can be only {\sc I-Sum},
{\sc Res}, {\sc Par}, {\sc Rep}, or {\sc Cong}. 
In the last four cases the 
assumption is again a $z w$ transition. 
By repeating this reasoning (descending the tree), 
we must arrive to a leaf transition of the form 
\[
P^I=\sum_i\alpha^I_i.Q_i \labtran{\alpha^I_j} Q_j\{w/v\}\;\;\mbox{ where }\alpha^I_j = z(v)
\]
Now, $P^O$ and $P^I$ must be two parallel 
processes in $P$, i.e. there must be a subprocess 
in $P$ of the form  $T[P^I] \,|\, U[P^O]$
(modulo $\equiv$), i.e. 
$P\equiv V[T[P^O] \,|\, U[P^I]]$
(here $T[\;]$, $U[\;]$ and $V[\;]$ represent  contexts, with the usual 
definition).
Furthermore, the $\bar{x}[y]$ transition
and the $z w$ transition must have been 
obtained by the application of the rule
{\sc Par} to this subprocess, i.e. 
$Q\equiv V[T[P_j]\,|\, U[P^I]]$ and 
$R\equiv V[T[P^O] \,|\, U[Q_j\{w/v\}]]$.
By applying again the rule {\sc Par} 
(plus all the other rules in the trees for the 
$\bar{x}[y]$ and the $z w$ transitions) 
we obtain 
the transitions 
$Q\labtran{z w} S$ 
and 
$Q'\labtran{\bar{x}[y]} S$ where $S= V[T[P_j]\,|\, U[Q_j\{w/v\}]]$.
\qed

\bigskip
\noindent

We are now ready to prove the  non-ex\-is\-tence result.
The intuition is the following: In the attempt to reach 
an agreement about the leader, the processes
of a symmetric network have to ``break the initial symmetry", 
and therefore have to communicate. 
The first such communication, however, can be repeated, 
by the above lemma and by symmetry, 
by all the pair of processes of the network. 
The result of all these transitions will still lead to a 
symmetric situation.
Thus there is a (infinite) computation in which the processes 
never succeed to break the symmetry, which means no leader 
is elected. 

\begin{theorem}\label{nonexistence api}
Consider a  network $P=[P_0\,|\,P_1\,|\ldots|\,P_{k-1}]$ in $\pi_s$, 
with $k \geq 2$.
Assume that $P$ has an 
automorphism $\sigma$ 
with only one orbit, 
and that $P$ is symmetric wrt $\sigma$.
Then $P$ cannot be an electoral system.
\end{theorem}

\noindent
{\bf Proof}
Assume by contradiction that $P$ is an electoral system. 
We will show that we can then construct an infinite 
increasing sequence
of computations for $P$, 
$C_0\prec C_1\prec\ldots \prec C_h\ldots$, 
such that for each $j$, $C_j:P\slabtran{\tilde\mu^j}P^j$ 
does not contain any  action 
of the form $\overline{\it out}\,n$, and $P^j$ is still symmetric wrt 
$\sigma_j$, where $\sigma_j$ is an automorphism 
with only one orbit obtained from $\sigma$ by adding associations on the 
new names introduced during the computation, 
and by eliminating the associations on the 
names that have disappeared.
This gives a contradiction, 
because the limit of this sequence is an infinite computation 
for $P$ which does not contain any action of the form $\overline{\it out}\,n$.

We will prove the above statement by induction. 
In order to understand the 
proof, it is important to notice that
if $\sigma$ has only one orbit then,
for each $i\in \{0,1,\ldots,k-1\}$, 
\[
{\it O}_{\sigma}(i)=
\{i,\sigma(i),\ldots ,\sigma^{k-1}(i)\}= \{0,1,\ldots,k-1\}
\]
\noindent
{\bf $h=0$)} 
Define $C_0$ to be the empty computation.

\noindent
{\bf $h+1$)}
Let  $C_h:P\slabtran{\tilde\mu^h}P^h$ where $P^h$
contains $k$ processes and
is symmetric wrt a
one-orbit automorphism $\sigma_h$.
We show how to  
construct 
$C_{h+1}:P\slabtran{\tilde\mu^{h+1}}P^{h+1}$, 
and a new one-orbit automorphism $\sigma_{h+1}$ for $P^{h+1}$,
such that $P^{h+1}$ contains $k$ processes and 
is symmetric wrt $\sigma_{h+1}$. 

Since $P$ is an electoral system, 
it must be possible to extend 
$C_h$ to a computation $C$ which contains ($k$) actions 
$\overline{\it out}\,n$, for a particular $n\in \{0,1,\ldots,k-1\}$.
Observe that the first action $\mu$ of 
$C\setminus C_h$ cannot be $\overline{\it out}\,n$. 
Otherwise, 
let $P^h_i$ be the component which performs 
this action. Then  $P^h_i$ must contain 
the subprocess $\overline{\it out}\,n$.
By symmetry, $P^h_{\sigma_h(i)}\equiv_\alpha \sigma_h(P^h_i)$ 
and therefore $P^h_{\sigma_h(i)}$ must contain 
the subprocess $\overline{\it out}\,\sigma(n)$. 
Furthermore, 
$\sigma_h(n)$ is free, because we have assumed 
that numbers cannot be used as bound names. 
Hence there must be an extension of $C$ 
where the action $\overline{\it out}\,\sigma_h(n)$ occurs. 
This implies (for the hypothesis that $P$ is an electoral system),
that $\sigma_h(n)=n$. Given that $k\geq 2$, 
$\sigma_h$ must generate more than one orbit. Contradiction.

In conclusion, $\mu$ must be an action different from 
$\overline{\it out}\,n$. We have two different situations depending on 
whether the transition is generated 
by the move of one process only, 
or by a communication between two processes 
(i.e. it involves the {\sc Com} or the {\sc Close} rule).
\begin{description}
\item[The transition is the result of the move of one process only)]
In this part of the proof, in 
order to simplify the notation we will assume, 
without loss of generality, that 
\[
\sigma(0) = 1, \;\sigma^2(0) = 2,\; \ldots 
\;\sigma^{k-1}(0)= k-1, \;\sigma^{k}(0) = 0
\] 
Therefore 
\[
P=[P_0\,|\,P_{\sigma(0)}\,|\ldots|\,P_{\sigma^{k-1}(0)}]
\]
 and, by symmetry, 
\[P\equiv_\alpha[P_0\,|\,\sigma(P_0)\,|\ldots|\,\sigma^{k-1}(P_{0})].\]

We also assume, without loss of generality, 
that $P^h_0$ is the component that 
performs the step, and let this step be
\[
P^h_0\;\;\labtran{\mu_0}\;\;P^{h+1}_0 
\]
\noindent
Using the symmetry of $P^h$, i.e. the fact that
$P^h_{i} = P^h_{\sigma_h^i(0)}
\equiv_\alpha 
\sigma_h^{i}(P^h_0)$ for every $i\in\{0,1,\ldots,k-1\}$, 
we can mimic the step of $P^h_0$ with every $P^h_{i}$ and 
derive the transitions
\[
\begin{array}{rcl}
P^h_1&\labtran{\mu_1}&P^{h+1}_1 \\
P^h_2&\labtran{\mu_2}&P^{h+1}_2 \\
&\vdots&\\
P^h_{k-1}&\labtran{\mu_{k-1}}&P^{h+1}_{k-1}
\end{array}
\]
such that 
\begin{itemize}
\item ${\mu_i}$ is not of the form 
$\overline{\it out}\,n$ for any $i,n\in\{0,1,\ldots,k-1\}$, 
\item the bound-names convention is respected in the $P^{h+1}_{i}$'s. 
\item the bound names of ${\mu_i}$ (if any) are different from the 
free names of all the other processes (and because of the bound-names 
convention in  $P^h$, we do not even
need to use $\alpha$ conversion here).
\end{itemize}
Thanks to the latter property, we can 
compose the displayed transitions into 
a computation
\[
P^h\;\;
\slabtran{\tilde\mu}\;\;
P^{h+1}
\]
where:
\begin{itemize}
\item $P^{h+1} = [P^{h+1}_0 \,\mid\, P^{h+1}_1\,\mid\ldots\mid\, P^{h+1}_{k-1}]$,
\item $\tilde{\mu}= \mu'_0 \mu'_1 \ldots \mu'_{k-1}$, 
where each $\mu'_i$ is equal to $\mu_i$
except if $\mu_i$
is a free output action, 
in which case the argument 
may become bound in $\mu'_i$ 
due to the restrictions at the top level in 
$P^h$.
\end{itemize}

It remains to show that we can 
construct a one-orbit automorphism $\sigma_{h+1}$ for 
$P^{h+1}$ such that $P^{h+1}$ is symmetric wrt it. 
To this purpose we need to distinguish various cases 
depending on $\mu_0$. 
\begin{description}
\item[$\mu_0 = \tau$)] 
In this case we have that, for every $i\in\{0,1,\ldots,k-1\}$, 
$\mu_i = \tau$ and 
$P^{h+1}_i \equiv_\alpha\sigma^i_h(P^{h+1}_0)$. 
Hence we can simply define $\sigma_{h+1}$ as $\sigma_h$ restricted to 
(the edges and the nodes of) $H(P^{h+1})$.

\vspace{.1in}
\item[$\mu_0 = \bar{x}_0 y_0$)]
We have that, for every $i\in\{0,1,\ldots,k-1\}$, 
$\mu_i = \overline{\sigma^i_{h}{\scriptstyle (}x_0{\scriptstyle )}} \sigma^i_{h}(y_0)$ and 
$P^{h+1}_i \equiv_\alpha\sigma^i_h(P^{h+1}_0)$. 
Hence also in this case we can define $\sigma_{h+1}$ as $\sigma_h$
restricted to $H(P^{h+1})$.

\vspace{.1in}
\item[$\mu_0 = \bar{x}_0 (y_0)$)]
We have that, for every $i\in\{0,1,\ldots,k-1\}$, 
$\mu_i = \overline{\sigma^i_{h}{\scriptstyle (}x_0{\scriptstyle )}} (y_i)$ 
for some $y_i$ that is different from every other bound and free 
name
in the other components of $P^{h}$.

Define $\sigma_{h+1}$ as follows: 
\begin{equation}\label{new automorphism}
\sigma_{h+1}(z) = \left\{ \begin{array}{ll}
                          y_{\sigma_h(i)}&\mbox{if}\; z = y_i 
				\mbox{ and } z  \in H(P^{h+1}) \\
                          \sigma_{h}(z)  &\mbox{if}\; z \neq y_i 
				\mbox{ and } z  \in H(P^{h+1})\\
                          \mbox{undefined} & \mbox{otherwise}
                          \end{array}
                  \right.
\end{equation}
Note that $\sigma_{h+1}$ is well-defined, because 
 for all $j\in\{0,1,\ldots,k-1\}$ \ $\sigma_{h+1}(j)= \sigma_{h}(j)$
since $j$ does not occur bound, and therefore $j$ cannot be one of the $y_i$'s.
Hence 
\[ 
P^{h+1}_{j} = P^{h+1}_{\sigma^j_{h+1}(0)}
\]
Finally, observe that 
$\sigma_{h+1}$ generates only one orbit and that
\[ P^{h+1}_{j}\equiv_\alpha \sigma^j_{h+1}(P^{h+1}_{0})\]
Therefore $P^{h+1}$ is symmetric wrt $\sigma_{h+1}$.

\vspace{.1in}
\item[$\mu_0 = x_0 y_0$)]
If $y_0$ is not new, i.e. it occurred in $P^{h}$, or if 
$y_0\in\{0,1,\ldots,k-1\}$, then 
we can choose the transitions above 
so that for every $i\in\{0,1,\ldots,k-1\}$, 
$\mu_i = \sigma^i_{h}(x_0)\sigma^i_{h}(y_0)$ and 
$P^{h+1}_i \equiv_\alpha\sigma^i_h(P^{h+1}_0)$. 
Hence also in this case we can define $\sigma_{h+1}$
as $\sigma_h$
restricted to $H(P^{h+1})$.

If, on the contrary, $y_0$ is new, then  
we can choose the transitions above 
so that for every $i\in\{0,1,\ldots,k-1\}$, 
$\mu_i = 
\sigma^i_{h}(x_0) y_i$ for some $y_i$ that is also new. 
(this is possible because in the early semantics we can 
instantiate an input action with an arbitrary name). 
Then, proceed as in the case ($\mu_0 = \bar{x}_0 (y_0)$). 

\end{description}

\item[The transition results from the communication of two processes)]
This is the part of the proof where we use the specific property of $\pi_s$
illustrated 
by Lemma \ref{confluence}.

The interesting case is when the 
two agents are in {\it different nodes} of the communication graph. 
(If the agents are inside the same node, say 
$P^h_i$, then we have a transition $P^h_i\;\;\labtran{\tau}\;\;P^{h+1}_i$
and we proceed 
like in previous case.)
Let $P^h_i$ and $P^h_j$ be the two processes, with $i\neq j$. 
We have two transitions 
$P^h_i\labtran{\mu_i}Q_i$ and $P^h_j\labtran{\mu_j}R_j$,
where $\mu_i$ and $\mu_j$ are complementary.
Assume without loss of generality that $\mu_i$ is the input action, and
$\mu_j$ is the output action. 
Since $\sigma_h$ generates only one orbit, there exists $r\in\{1,\ldots,k-1\}$
such that $j=\sigma_h^r(i)$. Assume 
for simplicity that $r$ and $k$ are relatively 
prime\footnote{If they are not, then in the rest of the proof 
$k$ has to be replaced by the least $p$ such that $pk=rq$, for some $q$.},
and let $\theta=\sigma_h^r$. Then  $P^h_{j}=P^h_{\theta(i)}$ and
we can write $R_{\theta(i)}$ for $R_j$. Let us first consider the 
case in which the first step of $C\setminus C_h$
has been produced by an application of the 
{\sc Com} rule. Then we have a transition
\[
P^h_i\mid P^h_{\theta(i)}\;\;\labtran{\tau}\;\;Q_i\mid R_{\theta(i)}
\]
By symmetry, we have that 
$P^h_{\theta(i)}\labtran{\theta(\mu_i)}\theta(Q_i)$.
By Lemma \ref{confluence}
we then have the transitions 
$R_{\theta(i)}\labtran{\theta(\mu_i)}R'$
and 
$\theta(Q_i)\labtran{\mu_j}R'$
for some $R'$. Let us define 
$P^{h+1}_{\theta(i)}=R'$.
By symmetry, we also have 
$P^h_{\theta^2(i)}\equiv P^h_{\theta(j)}\labtran{\theta(\mu_j)}\theta(R_{j})$, 
and $\theta(\mu_i)$, $\theta(\mu_j)$ are complementary, 
hence we can combine them into a transition 
\[
R_{\theta(i)}\mid P^h_{\theta^2(i)}\;\;\labtran{\tau}\;\; P^{h+1}_{\theta(i)}
\mid R_{\theta^2(i)}
\]
\noindent 
with $R_{\theta^2(i)}= \theta(R_{j})$. 
By repeatedly applying this reasoning, we obtain 
\[\begin{array}{rcl}
R_{\theta^2(i)}\mid P^h_{\theta^3(i)}&\labtran{\tau}& 
P^{h+1}_{\theta^2(i)}
\mid R_{\theta^3(i)}\\
&\vdots\\
R_{\theta^{k-2}(i)}\mid P^h_{\theta^{k-1}(i)}&\labtran{\tau}& 
P^{h+1}_{\theta^{k-2}(i)}
\mid R_{\theta^{k-1}(i)}
\end{array}
\]
\noindent 
and 
$R_{\theta^{k-1}(i)}\labtran{\theta^{k-1}(\mu_i)}P^{h+1}_{\theta^{k-1}(i)}$.
Finally, observe that from the transition 
$\theta(Q_i)\labtran{\mu_j}R'$
above we can derive 
$\theta^k(Q_i)\labtran{\theta^{k-1}(\mu_j)}\theta^{k-1}(R')$.
But $\theta^k=\sigma_h^{kr}={\it id}$, hence we have
$Q_i\labtran{\theta^k(\mu_j)}P^{h+1}_i$, 
where we have defined $P^{h+1}_i$ to be $\theta^{k-1}(R')$.
Therefore we can compose also these transitions, thus ``closing the 
circle'', and we obtain
\[
R_{\theta^{k-1}(i)}\mid Q_i\;\;\labtran{\tau}\;\; 
P^{h+1}_{\theta^{k-1}(i)}
\mid P^{h+1}_i 
\]
The composition of the displayed transitions 
gives us the intended continuation\footnote{Under the assumption that
$r$ and $k$ are relatively 
prime, also $\theta$ has only one orbit. 
If we drop this assumption, and hence we replace 
$k$ by the smallest $p$ such that $pk=rq$ for some $q$, 
then the computation we have constructed involves only the 
processes of the nodes in 
$O_\theta(i)=\{i,\theta(i),\ldots,\theta^{p-1}(i)\}$.
To complete computation we have to repeat the reasoning for 
the other orbits of $\theta$: $O_\theta(\sigma_h(i))$, 
$O_\theta(\sigma_h^2(i))$\ldots $O_\theta(\sigma_h^{q-1}(i))$.}:
\[
P^h = [P^{h}_i\,|\,P^{h}_{\theta(i)}\,|\ldots|\,P^{h}_{\theta^{k-1}(i)}]
\;\;\slabtran{\tilde{\tau}}\;\;
[P^{h+1}_i\,|\,P^{h+1}_{\theta(i)}\,|\ldots|\,P^{h+1}_{\theta^{k-1}(i)}]\footnote{We are using a sloppy notation here: 
the processes should be permuted so to have their indexes in increasing order.}
\]
\noindent
Finally define 
$P^{h+1}=
[P^{h+1}_i\,|\,P^{h+1}_{\theta(i)}\,|\ldots|\,
P^{h+1}_{\theta^{k-1}(i)}]$
and observe that $\sigma_h$ (restricted to $H(P^{h+1})$)
is still an automorphism for 
$P^{h+1}$ and that $P^{h+1}$ is still symmetric
with respect to it. Note that $H(P^{h+1})$ may differ 
from $H(P^{h})$ because some edges may have 
disappeared and because the {\sc Com} rule may have extended 
the set of nodes that share a certain edge. However, 
$H(P^{h+1})$ does not contain any new edges because {\sc Com} only 
transmits free names, corresponding to existing edges. 
Hence we can define $\sigma_{h+1}$ as the 
restriction of $\sigma_h$ to $H(P^{h+1})$. 

Consider now the case in which the 
first step of $C\setminus C_h$ 
is obtained by an application of the {\sc Close} rule. 
Then the transition is  of the form 
\[
\begin{array}{rcl}
P^h_i\mid P^h_{\theta(i)}&\labtran{\tau}&\nu y_i(Q_i\mid 
R_{\theta(i)}) 
\end{array}
\]
where $y_i$ is the name transmitted in the communication.
By following the same reasoning as before, we obtain the transitions
\[
\begin{array}{rcl}
R_{\theta(i)}\mid P^h_{\theta^2(i)}&\labtran{\tau}&
  \nu y_{\theta(i)}(P^{h+1}_{\theta(i)}\mid R_{\theta^2(i)})\\
R_{\theta^2(i)}\mid P^h_{\theta^3(i)}&\labtran{\tau}& 
  \nu y_{\theta^2(i)}(P^{h+1}_{\theta^2(i)}\mid R_{\theta^3(i)})\\
  &\vdots\\
R_{\theta^{k-2}(i)}\mid P^h_{\theta^{k-1}(i)}&\labtran{\tau}& 
  \nu y_{\theta^{k-2}(i)}(P^{h+1}_{\theta^{k-2}(i)}\mid R_{\theta^{k-1}(i)})\\
R_{\theta^{k-1}(i)}\mid Q_i&\labtran{\tau}& 
  \nu y_{\theta^{k-1}}(P^{h+1}_{\theta^{k-1}(i)}\mid P^{h+1}_i)
\end{array}
\]
Note that, thanks to the bound-names convention for 
$P^h$, all the $y_{j}$'s are different from each other and from 
the free variables. 
We can then combine the above transitions, and 
use the {\it Cong} rule with scope expansion
to push the restriction operators 
at the top-level of the network, 
thus obtaining the derivation
\[
P^h = [P^{h}_i\,|\,P^{h}_{\theta(i)}\,|\ldots|\,P^{h}_{\theta^{k-1}(i)}]
\;\;\slabtran{\tilde{\tau}}\;\;
[P^{h+1}_i\,|\,P^{h+1}_{\theta(i)}\,|\ldots|\,P^{h+1}_{\theta^{k-1}(i)}]
\]
Note that $H(P^{h+1})$  may contain some of the 
$y_j$'s as additional edges, because if these names occur 
in the $P^{h+1}_j$'s, 
they occur free. We need therefore to expand the automorphism
accordingly. This can be done by defining 
$\sigma_{h+1}$ exactly as in (\ref{new automorphism}). It is easy to see that
$\sigma_{h+1}$ has one orbit and that $P^{h+1}$ is symmetric wrt it.
\qed
\end{description}

\ \\

In \cite{Bouge:1988:AI} a less restrictive
 notion of symmetry is considered for proving negative results.
Namely, the automorphism $\sigma$ can have more orbits,
provided that they  all have the same cardinality (i.e. $\sigma$
can be well-balanced). 
In the framework of \cite{Bouge:1988:AI} this is a significant 
generalization, because the language considered there, 
$CSP_{\it in}$, can have the parallel operator only at the 
top level. Hence the condition of a single orbit, there, 
would impose that {\it all} the parallel processes present 
in the network have the same code (modulo renaming). 

In our framework, on the contrary, we do not have this restriction, 
and the above mentioned generalization is not essential. 
In fact, we can easily extend Theorem~\ref{nonexistence api} to well-balanced 
automorphisms:

\begin{corollary}\label{well balanced}
Consider a  network 
$P=[P_0\,|\,P_1\,|\ldots|\,P_{k-1}]$ in  $\pi_s$, 
and assume that the associated hypergraph 
$H(P)$ admits a well-balanced automorphism $\sigma\neq {\it id}$,
and that $P$ is symmetric wrt $\sigma$.
Then $P$ cannot be an electoral system.
\end{corollary}

\begin{figure}
\epsfbox{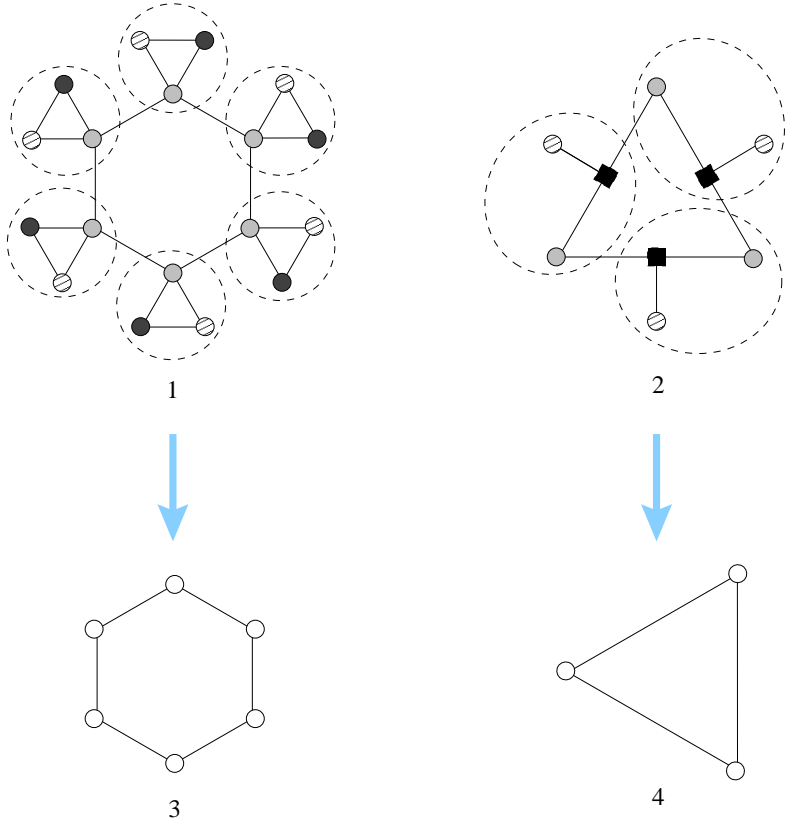}
\caption{Examples of hypergraphs with well-balanced automorphisms
and their transformation into hypergraphs with one-orbit 
automorphisms.}\label{well_balanced}
\end{figure}

Some examples of hypergraphs with well-balanced automorphisms are 
the hypergraphs $1$ and $2$ in Figure~\ref{well_balanced}. 
The nodes with the same filling represent nodes in the same orbit.

\ \\

\noindent
{\bf Proof of Corollary \ref{well balanced}}
The idea is to transform a network $P$ with a well-balanced 
automorphism into a network $Q$ with a one-orbit
automorphism by 
grouping together the nodes in the same orbit in $H(P)$ 
into one single node in $H(Q)$. 
For example, Hypergraphs $1$ and $2$ in Figure~\ref{well_balanced}
are transformed into Hypergraphs 
$3$ and $4$, respectively.

Assume that $\sigma$ generates $p$ orbits of cardinality $q$, and assume, 
without loss of generality, that $0,1,\ldots,p-1$ belong to different orbits,
and 
\[
\begin{array}{cccc}
\sigma(0) = p &\sigma(1) = p+1& \ldots &\sigma(p-1) = 2p-1\\
\sigma^2(0)= 2p& \sigma^2(1) = 2p+1& \ldots& \sigma^{2}(p-1) = 3p-1\\
\vdots&\vdots&\vdots&\vdots\\
 \ \ \sigma^{q-1}(0) = (q-1)p \ \ & \ \ \sigma^{q-1}(1) = (q-1)p+1 \ \  &  
\ \ \ldots \ \ &  \ \ \sigma^{q-1}(p-1) = qp-1 \ \ \\
\end{array}
\]
Define the processes
\[
\begin{array}{rcl}
Q_0 &=& P_{0}\,|\, P_{1}\,|\ldots |\, P_{p-1}\\
Q_1 &=& P_{p}\,|\, P_{p+1}\,|\ldots|\, P_{2p-1}\\
&\vdots&\\
Q_{q-1} &=& P_{(q-1)p}\,|\, P_{(q-1)p+1}\,|\ldots|\, P_{qp-1}
\end{array}
\]
Consider now the network 
$Q=[Q_0\,|\,Q_1\,|\ldots|\,Q_{q-1}]$. Clearly $Q$ and $P$ 
generate the same computations (they are strongly bisimilar), 
but the associated hypergraph, $H(Q)$, is different:
$H(Q)$ is ``an abstraction'' of  $H(P)$ in the sense that 
certain nodes of  $H(P)$ are ``grouped together'' in the 
same node of  $H(Q)$, as explained before. 

Note that $Q$ may contain names corresponding to non-existing nodes. 
To eliminate them, consider the renaming 
$\rho:\{0,1,\ldots,qp-1\}\rightarrow \{0,1,\ldots,q-1\}$ defined by 
\[
\rho(n) = n \;{\bf div}\; p
\]
i.e. $\rho(n)$ is the result of the integer division of $n$ by $p$, 
and define
$Q' = \rho(Q)$. 
It is easy to see that the traces of the projections of 
$Q'$ are the same as those in $Q$ modulo the renaming $\rho$.

In the hypergraph $H(Q')$ the nodes $N$ are, of course, $0,1,\ldots,p-1$. 
The edges $X$ are the same as the 
edges of $H(P)$ minus $\{p,p+1,\ldots,qp-1\}$. 
(Note that in Figure~\ref{well_balanced} the edges internal 
to the nodes of the transformed graph are not represented.) 

Now, consider the pair $\theta=\la \theta_N,\theta_X\ra$ 
with $\theta_N(0)=1$, $\theta_N(1)=2$,\ldots, $\theta_N(q-1)=0$,
and  $\theta_X=\sigma_X$ restricted to $H(Q')$. It is easy to see 
that $\theta$ is an automorphism on $H(Q')$ with only one orbit, 
and that $Q'$ is symmetric wrt $\theta$.

Finally, observe that if $P$ is an electoral system then also
$Q'$ is an electoral system, and apply Theorem~\ref{nonexistence api}.
\qed

\section{Existence of symmetric electoral systems in $\pi_m$}\label{pi_m versus pi_I and CCS}
The negative result of previous section does not apply to $\pi_m$:
its mixed-choice construct, in fact, makes it possible to 
establish a simultaneous a\-gree\-ment among 
two processes, thus breaking the symmetry. 
Note that the presence of mixed choice invalidates the confluence 
property of Lemma~\ref{confluence}.

Consider for example the election problem 
in a symmetric network consisting of two nodes $P_0$ and $P_1$
only, and two private edges, $x_0$ and $x_1$, connecting them. 
A $\pi$-calculus  specification which solves the problem is
$P = \nu x_0\, \nu x_1\ (P_0\,|\,P_1)$, where:
\begin{equation}\label{two-nodes system}
\renewcommand{\arraystretch}{1}
\begin{array}{rcl}
P_i&= &\overline{x}_i y.\overline{\it out}\ i\\
   &  &+\\
   &  &{x_{i\oplus 1}}(y).\overline{\it out}\ \langle i\oplus 1\rangle
\end{array}
\renewcommand{\arraystretch}{1.5}
\end{equation}
\noindent
with $i\in\{0,1\}$ and $\oplus$ being the sum modulo $2$. The argument $y$ is not
relevant, it is present only because in $\pi_m$ actions must have an argument.

Note that the above electoral system, although very simple, 
has an automorphism with only one orbit, 
hence by Theorem~\ref{nonexistence api} it cannot be 
expressed in $\pi_s$. 

What happens when the hypergraph is  more complicated?
We claim that in $\pi_m$ the existence of  
symmetric electoral systems is guaranteed 
in a large number of cases: 

\begin{claim}\label{existence pi}
Let $H$ be a {\it connected} hypergraph 
(i.e. each pair of nodes are connected 
by a sequence of edges).
Then there exists a symmetric electoral system 
$P$ in $\pi_m$ such that 
$H(P)=H$. (Remember that ``symmetric'' 
means symmetric wrt every possible automorphism 
on the hypergraph.)
\end{claim}

Our claim is substantiated by the following 
idea for an electoral algorithm:

Let $k$ be the number of nodes.
The generic process
$P_i$ executes the following: 
\begin{enumerate}
\item Broadcast a private name $y_i$ 
to all the other processes (possible thanks to the connectivity hypothesis) 
and, meanwhile, 
receive the private name $y_j$ of 
each other process $P_j$. In this way the hypergraph becomes fully connected.
\item Repeat (at most $k$ times) a choice where one guard
is an output action on $y_i$, while the 
others are input actions on the  $y_j$'s.
If at a certain point an 
input is selected, then go to  4.
\item If this point is reached, then $P_i$ is the 
leader. Broadcast this  information to all the other processes, 
output $\overline{\it out}\ i$ and terminate.
\item Wait to receive the name of the leader. 
Then send it on {\it out} and terminate.
\qed
\end{enumerate}

Note that the above algorithm works under the assumption  that each process
knows what is the total number of processes in the network. 

It is difficult to make the above 
argument more formal while keeping it 
general, since 
the details of the algorithm 
(like how to broadcast the private name) 
depend on the structure of the hypergraph. 

\begin{figure}
\setlength{\unitlength}{0.00045in}
\begingroup\makeatletter\ifx\SetFigFont\undefined%
\gdef\SetFigFont#1#2#3#4#5{%
  \reset@font\fontsize{#1}{#2pt}%
  \fontfamily{#3}\fontseries{#4}\fontshape{#5}%
  \selectfont}%
\fi\endgroup%
{\renewcommand{\dashlinestretch}{30}
\begin{picture}(3836,3585)(0,-10)
\path(472,288)(3345,288)(3345,3162)
	(472,3162)(472,288)
\put(3531,3300){\makebox(0,0)[lb]{\smash{{{\SetFigFont{12}{14.4}{\rmdefault}{\mddefault}{\updefault}2}}}}}
\put(100,3300){\makebox(0,0)[lb]{\smash{{{\SetFigFont{12}{14.4}{\rmdefault}{\mddefault}{\updefault}3}}}}}
\put(3531,0){\makebox(0,0)[lb]{\smash{{{\SetFigFont{12}{14.4}{\rmdefault}{\mddefault}{\updefault}1}}}}}
\put(100,0){\makebox(0,0)[lb]{\smash{{{\SetFigFont{12}{14.4}{\rmdefault}{\mddefault}{\updefault}0}}}}}
\put(1900,0){\makebox(0,0)[b]{\smash{{{\SetFigFont{12}{14.4}{\rmdefault}{\mddefault}{\updefault}$x_0$}}}}}
\put(1900,3375){\makebox(0,0)[b]{\smash{{{\SetFigFont{12}{14.4}{\rmdefault}{\mddefault}{\updefault}$x_2$}}}}}
\put(231,1650){\makebox(0,0)[b]{\smash{{{\SetFigFont{12}{14.4}{\rmdefault}{\mddefault}{\updefault}$x_3$}}}}}
\put(3606,1650){\makebox(0,0)[b]{\smash{{{\SetFigFont{12}{14.4}{\rmdefault}{\mddefault}{\updefault}$x_1$}}}}}
\put(472,288){\whiten\ellipse{274}{274}}
\put(472,288){\ellipse{274}{274}}
\put(3345,3162){\whiten\ellipse{274}{274}}
\put(3345,3162){\ellipse{274}{274}}
\texture{88555555 55000000 555555 55000000 555555 55000000 555555 55000000 
	555555 55000000 555555 55000000 555555 55000000 555555 55000000 
	555555 55000000 555555 55000000 555555 55000000 555555 55000000 
	555555 55000000 555555 55000000 555555 55000000 555555 55000000 }
\put(472,3162){\shade\ellipse{274}{274}}
\put(472,3162){\ellipse{274}{274}}
\put(3345,288){\shade\ellipse{274}{274}}
\put(3345,288){\ellipse{274}{274}}
\end{picture}
}
\caption{A hypergraph which is connected, but not fully connected
since the two pairs of nodes with the same fillings are not directly 
connected by any edge.}\label{hquadrato}
\end{figure}
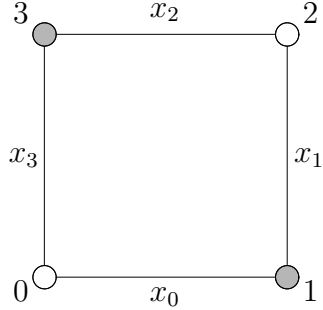

For proving separation results between $\pi_m$ 
and the other languages, 
however, it is sufficient 
to show that $\pi_m$ can solve the symmetric electoral problem 
in one hypergraph, suitably chosen. 
We will consider the simple hypergraph 
in Figure~\ref{hquadrato}.

\begin{proposition}\label{quadrato}
Let $H$ be the hypergraph illustrated in Figure~\ref{hquadrato}.
Then, there exists a symmetric electoral system 
$P = \nu x_0\, \nu x_1\, \nu x_2\, \nu x_3\, (P_0\, |\, P_1\, |\, P_2\, |\, P_3)$, in $\pi_m$, such that 
$H(P)=H$. 
\end{proposition}

\noindent
{\bf Proof}
For every  $i\in \{0,1,2,3\}$ we define $P_i$ following the idea
illustrated above. First $P_i$ broadcasts its private name $y_i$ 
to all the other nodes, and receives the private names of all the 
other nodes (first phase). Then each process uses $y_i$ to compete 
for the election (second phase). 

For the first phase of the algorithm, we define 
\begin{equation}\label{algorithm_first}
 P_i = \nu y_i\,  P_i^3
\end{equation}
where, intuitively, 
$P_i^3$ 
represents a process that 
receives three names from its left neighbor via the channel $x_{i-i}$
and sends three names to its right neighbor via the channel $x_i$, in such a way that
the first name sent is $y_i$, and
the other two are the first two names received from the left neighbor. In this way 
each name is broadcasted to all processes.

At this point it would seem natural 
to define  
$P_i^k = \overline{x}_{i} y_{i+k+1} \,|\,x_{i-1}(y_{i+k})\,.\, P_i^{k-1}$
for $1\leq k\leq 3$, but there is a problem: the output 
$\overline{x}_{i} y_{i+k}$ in $P_i^{k-1}$
may become available before 
$\overline{x}_{i} y_{i+k+1}$ 
is consumed, so the right neighbor may receive 
$y_{i+k}$ before $y_{i+k+1}$. 
This would be incorrect because the order is relevant: 
the first name received 
(intended to be $y_{i+k+1}$) 
will be 
retransmitted 
$k-1$ times while the second (intended to be $y_{i+k}$)
will be retransmitted only $k-2$ times. 

In order to solve the above problem we need to sequentialize the output actions. 
We do this by using an idea similar 
the one used Honda and Tokoro (1991) for the encoding of the output prefix. 
Namely, we replace each action by  
a pair of actions:  $\overline{x}y$ becomes $x(w)$ followed by $\overline{w}y$, 
and $x(y)$ becomes $\overline{x}w$, where $w$ is a fresh name, and $w(y)$. 

The definition of $P_i^k$ then is as follows 
(the symbols $+$ and $-$ here 
represent the sum and difference modulo $4$, 
respectively): 

\begin{equation}\label{algorithm}
\begin{array}{rcll}
P_i^k & = & \nu w\ (\overline{x}_{i-1} w \ | \ x_i(w')\ .\ (\overline{w}'y_{i+k+1} \ | \ w(y_{i+k})\ .\ P_i^{k-1})\ )  
    &\mbox{for } 1 \leq k \leq 3\\
P_i^0 & = &Q^3_i 
\end{array}
\end{equation}

\begin{figure}
\setlength{\unitlength}{0.00051667in}
\begingroup\makeatletter\ifx\SetFigFont\undefined%
\gdef\SetFigFont#1#2#3#4#5{%
  \reset@font\fontsize{#1}{#2pt}%
  \fontfamily{#3}\fontseries{#4}\fontshape{#5}%
  \selectfont}%
\fi\endgroup%
{\renewcommand{\dashlinestretch}{30}
\begin{picture}(3836,3585)(0,-10)
\put(3821.000,3543.000){\arc{6775.743}{2.7016}{3.0253}}
\put(1356.000,4264.000){\arc{4491.279}{0.5190}{1.8413}}
\put(6016.000,6846.000){\arc{14169.279}{1.9633}{2.4075}}
\put(-2844.000,1775.000){\arc{7229.281}{6.1932}{6.7035}}
\put(3699.000,-215.000){\arc{6775.743}{4.2724}{4.5961}}
\put(4420.000,2250.000){\arc{4491.279}{2.0898}{3.4121}}
\put(7002.000,-2410.000){\arc{14169.279}{3.5341}{3.9783}}
\put(1931.000,6450.000){\arc{7229.281}{1.4808}{1.9911}}
\put(-59.000,-93.000){\arc{6775.743}{5.8432}{6.1669}}
\put(2406.000,-814.000){\arc{4491.279}{3.6606}{4.9829}}
\put(-2254.714,-3396.000){\arc{14170.339}{5.1050}{5.5492}}
\put(6606.000,1675.000){\arc{7229.281}{3.0516}{3.5619}}
\put(63.000,3665.000){\arc{6775.743}{1.1308}{1.4545}}
\put(-658.286,1200.000){\arc{4491.830}{5.2315}{6.5536}}
\put(-3240.429,5860.714){\arc{14170.914}{0.3926}{0.8367}}
\put(1831.000,-3000.000){\arc{7229.281}{4.6224}{5.1327}}
\blacken\path(831,2025)(831,2175)(681,2175)
	(681,2025)(831,2025)
\path(831,2025)(831,2175)(681,2175)
	(681,2025)(831,2025)
\put(3491,0){\makebox(0,0)[lb]{\smash{{{\SetFigFont{12}{14.4}{\rmdefault}{\mddefault}{\updefault}1}}}}}
\put(906,2325){\makebox(0,0)[b]{\smash{{{\SetFigFont{12}{14.4}{\rmdefault}{\mddefault}{\updefault}$y_3$}}}}}
\put(2481,2600){\makebox(0,0)[b]{\smash{{{\SetFigFont{12}{14.4}{\rmdefault}{\mddefault}{\updefault}$y_2$}}}}}
\put(2901,1050){\makebox(0,0)[b]{\smash{{{\SetFigFont{12}{14.4}{\rmdefault}{\mddefault}{\updefault}$y_1$}}}}}
\blacken\path(2181,2775)(2331,2775)(2331,2925)
	(2181,2925)(2181,2775)
\path(2181,2775)(2331,2775)(2331,2925)
	(2181,2925)(2181,2775)
\blacken\path(2931,1425)(2931,1275)(3081,1275)
	(3081,1425)(2931,1425)
\path(2931,1425)(2931,1275)(3081,1275)
	(3081,1425)(2931,1425)
\blacken\path(1581,675)(1431,675)(1431,525)
	(1581,525)(1581,675)
\path(1581,675)(1431,675)(1431,525)
	(1581,525)(1581,675)
\dashline{60.000}(3306,300)(3306,3150)(456,3150)
	(456,300)(3306,300)
\put(3491,3300){\makebox(0,0)[lb]{\smash{{{\SetFigFont{12}{14.4}{\rmdefault}{\mddefault}{\updefault}2}}}}}
\put(101,3300){\makebox(0,0)[lb]{\smash{{{\SetFigFont{12}{14.4}{\rmdefault}{\mddefault}{\updefault}3}}}}}
\put(101,0){\makebox(0,0)[lb]{\smash{{{\SetFigFont{12}{14.4}{\rmdefault}{\mddefault}{\updefault}0}}}}}
\put(1881,50){\makebox(0,0)[b]{\smash{{{\SetFigFont{12}{14.4}{\rmdefault}{\mddefault}{\updefault}$x_0$}}}}}
\put(1881,3325){\makebox(0,0)[b]{\smash{{{\SetFigFont{12}{14.4}{\rmdefault}{\mddefault}{\updefault}$x_2$}}}}}
\put(231,1650){\makebox(0,0)[b]{\smash{{{\SetFigFont{12}{14.4}{\rmdefault}{\mddefault}{\updefault}$x_3$}}}}}
\put(3556,1650){\makebox(0,0)[b]{\smash{{{\SetFigFont{12}{14.4}{\rmdefault}{\mddefault}{\updefault}$x_1$}}}}}
\put(1301,750){\makebox(0,0)[b]{\smash{{{\SetFigFont{12}{14.4}{\rmdefault}{\mddefault}{\updefault}$y_0$}}}}}
\texture{88555555 55000000 555555 55000000 555555 55000000 555555 55000000 
	555555 55000000 555555 55000000 555555 55000000 555555 55000000 
	555555 55000000 555555 55000000 555555 55000000 555555 55000000 
	555555 55000000 555555 55000000 555555 55000000 555555 55000000 }
\put(457,3167){\shade\ellipse{274}{274}}
\put(457,3167){\ellipse{274}{274}}
\put(455,317){\whiten\ellipse{274}{274}}
\put(455,317){\ellipse{274}{274}}
\put(3289,300){\shade\ellipse{274}{274}}
\put(3289,300){\ellipse{274}{274}}
\put(3289,3135){\whiten\ellipse{274}{274}}
\put(3289,3135){\ellipse{274}{274}}
\end{picture}
}
\caption{The evolution of the hypergraph in Figure~\ref{hquadrato} 
after running the first phase of the algorithm, i.e.the code
 described in (\ref{algorithm_first}) and (\ref{algorithm}) up to the $Q^3_i$'s.}
\label{hquadrato_comp}
\end{figure}
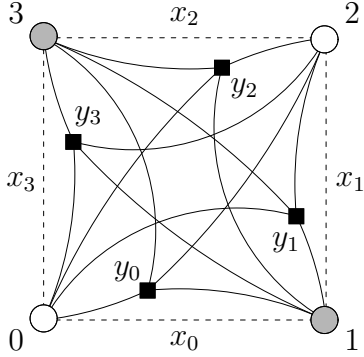

The purpose of $Q^3_i$ is to perform the  second phase 
of the algorithm, i.e. to compete for the election.
The meaning of the superscript $3$ will be clear later. 
By the time each $P_i$ reaches the point  $Q^3_i$, 
the original hypergraph has evolved into the 
fully connected hypergraph
illustrated in Figure~\ref{hquadrato_comp}. 
Note that the $x_i$'s are no longer there because 
they do not occur in the $Q^3_i$'s.

Interestingly, this first phase of the algorithm 
is within the syntax of 
$\pi_s$ and $\pi_a$, but not of  $\pi_{I}$
nor of the $\pi$-calculus version of CCS$_{vp}$. 
By ``$\pi$-calculus version of CCS$_{vp}$''
we mean the language obtained from CCS$_{vp}$ by replacing the
restriction operator of CCS$_{vp}$ with the one of the $\pi$-calculus, 
and other obvious changes of this kind.

We now describe the $Q^k_i$'s, for $0 \leq k \leq 3$.
As stated above, their purpose is to compete for the election.
Each node $i$ tries to ``dominate'' the other 
nodes by sending messages on its own channel $y_i$. 
Each time $i$ sends a message on $y_i$ to another node, say $j$, 
the nodes dominated by $j$, and $j$ itself, become dominated 
by $i$. The superscript 
$k$, initially $3$, in $Q^k_i$ 
indicates how many additional nodes $i$ needs to dominate 
before becoming a winner.
Obviously $i$ becomes the winner whenever 
it succeed to dominate all the other nodes, 
i.e. whenever $k=0$.
At this point $i$ executes $Q^0_i$, namely it tells the 
name of the winner (its own name $i$) to
each other node $j$ by using $j$'s channel $y_j$, 
and then it outputs its own 
name $i$ on {\it out}. 
Note that in order for $i$ to know that it is the winner 
it is sufficient to keep track of the {\it number} of nodes that it 
dominates\footnote{In a 
preliminary version of this paper 
there was an error: the number of nodes currently 
dominated by $i$ was calculated by adding 
$1$ each time $i$ was doing an output on $y_i$. 
The error was pointed out by Peng Wu, 
and corrected by the author in the way described here. Note that the 
corresponding code in (\ref{alg_II}) is tuned for the case of $4$ nodes.
Later Peng Wu has coded the algorithm for the case of arbitrary rings,
by using counters, and he has verified his program under the Mobility 
Workbench model checker.}. 
In order to do this, whenever $i$ sends a message on $y_i$ to a node $j$,
the message is a new name $z$. Then $j$ sends  back to $i$ a new name $s$. 
These names $z$ and $s$ are used by $j$ to communicate to $i$ the number $d$
of nodes that it dominates: 
$j$ does this by outputting $d$ times on $s$ and then one time on $z$.
Correspondingly, $i$ executes the part $R^k_i$, 
namely it receives all the ($d$)
inputs on $s$, 
then one input on $z$, and finally it becomes $Q^{k-d-1}_i$. 
Note that $R^0$ is never reached, so it does not need to be defined.

Of course, by symmetry, we need to make 
it possible for $i$ to lose the competition. Thus in alternative to 
the output on $y_i$ the node $i$ must also try to receive input from all the 
other nodes, on their own channels. If one of these input 
actions succeeds, then $i$ becomes dominated and it exits the competition.
At this point $i$ executes $S_i^{d}$:
first it comunicates to the dominator a new name $s$ (whose purpose is described above)
and the number ($3-k$) of nodes that $i$ 
currently dominates, 
and then (last two instructions of 
$S_i^d$) it waits to receive the name of the winner 
(from the winner), on its own channel, and then it sends such name on 
{\it out}.

The definition of the $Q^k_i$'s is as follows:
\begin{equation}\label{alg_II}
\renewcommand{\arraystretch}{1}
\begin{array}{rcll}
Q^k_i&= &\nu z
         \begin{array}[t]{ll}
         (
         &\overline{y}_{i}z\ . \ z(s)\ . \ R^{k}_i\ \\
         &+\\
         &y_{i+1}(z)\ . \ S^{3-k}_i \\
         &+\\
         &y_{i+2}(z)\ . \ S^{3-k}_i\\
         &+\\
         &y_{i+3}(z)\ . \ S^{3-k}_i\\
         )
         \end{array}
         &\mbox{for } 1 \leq k \leq 3     
\\
\ \\
Q^0_i&= &\overline{y}_{i+1}i\ . \ \overline{y}_{i+2}i  \ . \ \overline{y}_{i+3}i  \ . \  \overline{\it out}\ i
\\
\ \\
R^k_i &= & \begin{array}[t]{l}
         s(w)\ . \ R^{k-1}_i\\
         +\\
         z(w)\ . \ Q^{k-1}_i
         \end{array}
\\
\ \\
S^d_i &= & \nu s \ (\ \overline{z}s\ . \ \underbrace{\overline{s}s\ . \ \ldots \ . \ 
\overline{s}s}_{\mbox{$d$ times}}\ . \ \overline{z}s\ . \ y_{i}(n)\ . \ \overline{\it out}\ n\ )
\end{array}
\renewcommand{\arraystretch}{1.5}
\end{equation}

Note that the $P_i$'s have all the same code except for a renaming which 
corresponds to the structure of the hypergraph, hence $P$ is symmetric wrt every 
automorphism on the hypergraph.
\qed

\ \\ \ 

In contrast to the first phase of the 
algorithm, the second phase, displayed in (\ref{alg_II}), is not within the syntax
of $\pi_a$ or $\pi_s$, but it is within the syntax of $\pi_I$, and
an analogous process can also be written in CCS$_{vp}$. 
Hence for an hypergraph like the one in Figure~\ref{hquadrato_comp}
the symmetric electoral problem can be solved also in these two languages. 
More in general, we believe that the problem can be solved 
in CCS$_{vp}$ or in
$\pi_{I}$ for {\it any fully connected graph}.

\begin{claim}
Let $H$ be a {\it fully connected} hypergraph 
(i.e. each pair of nodes are connected 
directly by an edge).
Then there exists a symmetric electoral system 
$P$, in CCS$_{vp}$ or in $\pi_I$, such that 
$H(P)=H$. 
\end{claim}

Again, it is difficult to prove this claim in general (wrt any fully connected 
hypergraph)
because the precise steps of the algorithm depend on the structure of the 
hypergraph.

In the next section, we investigate the limitations of CCS$_{vp}$ and 
$\pi_{I}$ by showing a class of hypergraphs for which the 
symmetric electoral problem cannot be solved with these two languages. 

\section{Non existence of symmetric electoral systems in CCS$_{vp}$ and in 
$\pi_{I}$}
The mechanisms of name-passing and scope extrusion, which makes it possible 
in the $\pi$-calculus (and in $\pi_m$) to extend dynamically the communication 
structure of the network, 
are essential for the solution to the electoral problem 
illustrated in Proposition~\ref{quadrato}. 
In fact, as we shall see, 
the part of the solution which 
completes the hypergraph cannot be expressed  
in CCS$_{vp}$ or in $\pi_{I}$.

More in general, CCS$_{vp}$ and $\pi_{I}$ cannot express any solution to 
the symmetric electoral problem in a hypergraph like the one of 
Figure~\ref{hquadrato}. 
Intuitively, the problem is that
the symmetry cannot be broken as long as there is no direct connection
(channel)  between symmetric nodes (i.e. the nodes in the same orbit).
And in CCS$_{vp}$, as well as in $\pi_{I}$, there is no way to create a new 
direct connection between two nodes, unless, in the case of 
$\pi_{I}$, they are already sharing a channel. 

\begin{theorem}\label{nonexistence CCS}
Let $P=[P_0\,|\,P_1\,|\ldots|\,P_{k-1}]$
be a  network in CCS$_{vp}$ or in $\pi_{I}$,
and let the associated hypergraph $H(P)=\la N, X,t\ra$
admit a
well-balanced automorphism $\sigma\neq {\it id}$ such that 
$P$ is symmetric wrt $\sigma$ and, for each $n\in N$, 
there exist no $h$ such that 
$\sigma^h(n)\neq n$ and 
$\{n,\sigma^h(n)\}\subseteq t(x)$
for some $x\in X$. 
Then $P$ cannot be an electoral system. 
\end{theorem}

An example of such network is represented in Figure~\ref{hquadrato}: 
let $\sigma$ be the automorphism defined as $\sigma(0) = 2$,  $\sigma(2) = 0$,
$\sigma(1) = 3$, and $\sigma(3) = 1$. Clearly $\sigma$ is well balanced (it
has two orbits of cardinality two).  Note that the hypotheses of the above theorem
are satisfied, because there is no edge between $0$ and $2$, and between 
$1$ and $3$. 

\ \\

\noindent
{\bf Proof of Theorem \ref{nonexistence CCS}} 
The proof is analogous to that of Theorem~\ref{nonexistence api}, and it is based on the 
construction of an infinite computation from $P$ where no leader is elected.

Suppose that at a certain step of the computation $P$ has evolved into 
$P^h$, no leader has been elected yet, 
and $P^h$ is symmetric wrt a well balanced automorphism 
$\sigma_h$ which satisfies the hypotheses of the theorem. 
Consider the first step from $P^h$. For the same reasons
explained in the proof of Theorem~\ref{nonexistence api}, 
this step cannot be of the form $\overline{\it out}\,n$.
We have two cases: 
\begin{description}

\item[The transition is the result of the move of one process only)]
Assume, without loss of generality, that $P^h_0$ 
is the process which makes the move. 
Let this move be
\[P^h_0\;\labtran{\mu_{0}}\;P^{h+1}_0\]
Since $P^h_{\sigma^i_h(0)} \equiv_\alpha \sigma^i_h(P^h_0)$, 
By symmetry, we can construct derivations
\[
\begin{array}{rcl}
P^h_{\sigma_h(0)} &\labtran{\mu_{1}}& P^{h+1}_{\sigma_h(0)}\\
P^h_{\sigma^2_h(0)} &\labtran{\mu_{2}}& P^{h+1}_{\sigma^2_h(0)}\\
&\vdots\\
P^h_{\sigma^{q-1}_h(0)} &\labtran{\mu_{q-1}}& P^{h+1}_{\sigma^{q-1}_h(0)}
\end{array}
\]
(where $q$ is the cardinality of the orbits of $\sigma$) such that 
\[
P^h_{\sigma^i_{h+1}(0)} \equiv_\alpha \sigma^i_{h+1}(P^h_0)
\]
where $\sigma_{h+1}$ is constructed by adding to $\sigma_h$ the associations 
on the new names introduced by the above transitions, if any, as in 
the proof of Theorem~\ref{nonexistence api}. 
Note that $\sigma_{h+1}$ is well balanced and coincides with $\sigma_{h}$ 
on all the processes $P^{h}_j$ such that $j$ not in the orbit of $0$. 
For such processes, define  $P^{h+1}_j=P^{h}_j$. 
By composing the transitions above, we get a non-empty 
sequence of transitions from $P^{h}$ to 
$P^{h+1}=[P^{h+1}_0\,|\,P^{h+1}_1\,|\ldots|P^{h+1}_{k-1}]$
which does not contain $\overline{\it out}\,n$. Finally, observe that 
$P^{h+1}$ is symmetric wrt $\sigma_{h+1}$.

\vspace{.1in}
\item[The transition results from the communication of two processes)]
This is the crucial part of the proof, which distinguishes 
CCS$_{vp}$ and  $\pi_{I}$ from $\pi_m$. Again, the interesting case is when 
the two communicating processes are in different nodes.
Assume, without loss of generality, that $P^h_0$ and $P^h_i$ are the partners in
the communication. 
Note that, by the hypothesis that processes in the same orbit are not connected, 
$0$ and
$i$ must be in different orbits. 
Let us consider the case of $\pi_{I}$ first. We have 
that the communication can only derive from the rule {\sc Close}, because 
output actions in $\pi_{I}$ can only be of the form $\bar{x}(y)$. Assume without loss
of generality that $P^h_0$ is the sender and that the communication action is 
$\bar{x}_0(y_0)$. we have:
\[
\begin{array}{lcl}
P^h_0|P^h_i&\labtran{\tau}& \nu y_0\ (P^{h+1}_0|P^{h+1}_i)
\end{array}
\]
By symmetry, i.e. since 
$P^h_{\sigma^j_h(0)} \equiv_\alpha \sigma^j_h(P^h_{0})$ and
$P^h_{\sigma^j_h(i)} \equiv_\alpha \sigma^j_h(P^h_{i})$, 
we also have
\[
\begin{array}{rcl}
P^h_{\sigma_h(0)}|P^h_{\sigma_h(i)}&\labtran{\tau}& \nu y_{\sigma_{h(0)}}\ (P^{h+1}_{\sigma_h(0)}|P^{h+1}_{\sigma_h(i)})\\
P^h_{\sigma^2_h(0)}|P^h_{\sigma^2_h(i)}&\labtran{\tau}& \nu y_{\sigma^2_h(0)}\ (P^{h+1}_{\sigma^2_h(0)}|P^{h+1}_{\sigma^2_h(i)})\\
&\vdots\\
P^h_{\sigma^{q-1}_h(0)}|P^h_{\sigma^{q-1}_h(i)}&\labtran{\tau}& \nu y_{\sigma^{q-1}_h(0)}(P^{h+1}_{\sigma^{q-1}_h(0)}|P^{h+1}_{\sigma^{q-1}_h(i)})\\
\end{array}
\]
where all $y_l$'s are distinct, and distinct from the free names 
(by the bound-names convention on $P^h$), and 
\[ 
P^{h+1}_{\sigma^j_{h+1}(0)} \equiv_\alpha \sigma^j_{h+1}(P^{h+1}_{0}) \;\mbox{ and }\;
P^{h+1}_{\sigma^j_{h+1}(i)} \equiv_\alpha \sigma^j_{h+1}(P^{h+1}_{i})
\] 
where $\sigma_{h+1}$ is defined as in (\ref{new automorphism}).
For any $j$ which is neither in the orbit of $0$, nor in the orbit of $i$, define
$P^{h+1}_j=P^{h}_j$, and let
\[P^{h+1}=[\nu y_0\ \nu y_1 \ldots\nu y_{q-1} (P^{h+1}_0\,|\,P^{h+1}_1\,|\ldots|P^{h+1}_{k-1})]\] 
By using the {\it Cong} rule with scope expansion, 
we can combine the above transitions into a computation
\[
P^h \;\;\slabtran{\tilde{\tau}}\;\;P^{k+1}
\]
Finally, note that:
\begin{itemize}
\item $P^{k+1}$ is symmetric wrt to $\sigma_{h+1}$, and
\item $H(P^{k+1})$ differs from $H(P^{k})$ only for the presence of the new edges
$y_j$ between the nodes $P^{h+1}_{\sigma^j_h(0)}$ and $P^{h+1}_{\sigma^j_h(i)}$
(which were already connected by the  edge $x_{\sigma^j_h(0)}$). 
None of the existing edges have changed their type, i.e. two nodes that 
were not connected by any edge in $H(P^{h})$ are still not
connected by any edge in  $H(P^{h+1})$.
\end{itemize}

In the case of CCS$_{vp}$, the proof is analogous. The crucial point here is that 
the objects of the communications, i.e. the $y_j$'s, can only be values. 
Therefore they cannot be used as communication channels in later steps of the 
computation\footnote{In the case of CCS$_{vp}$ the notion of hypergraph associated 
to a network should be slightly different, i.e. we should distinguish between 
edges that represent channels and edges that represent only values and 
therefore cannot be used as communication channels. Also, we should use scope 
extrusion as a bisimilarity law.}.
\qed
\end{description}

\section{Uniform encoding}\label{uniform encoding}
In this section we use the above results 
to show the non-encodability of 
the $\pi_m$ into its asynchronous subsets, into 
CCS$_{vp}$, and into $\pi_I$, under certain requirements on the 
notion of encoding  $\os\cdot\cs$.

There is no agreement on what should be a good notion of encoding, 
and perhaps indeed there should not be a unique notion, but several, 
depending on the purpose. 
However, it seems reasonable to require at least the two following 
properties:
\begin{enumerate}
\item compositionality,
\item preservation of some intended semantics. 
\end{enumerate}

For a distributed system, however, it seems reasonable to 
strengthen the notion of compositionality on the parallel operator
by requiring 
that  it is translated homomorphically, namely
\begin{eqnarray}\label{parallel pres}
\os P\,|\,Q\cs &=& \os P\cs \;|\;\os Q\cs
\end{eqnarray}
In this way we can ensure that the
translation 
maintains the degree of distribution of the system, without introducing 
additional processes with coordination functions.

Likewise, it seems reasonable to require that the encoding 
``behaves well'' 
with respect to channel renamings, i.e. for any permutation of 
names $\sigma$ in the domain of the source language there exists a
 permutation of 
names $\theta$ in the domain of the target language such that 
$\forall i\in N\; \sigma(i) = \theta(i)$ and 
\begin{eqnarray}\label{renaming pres}
\os \sigma(P) \cs = \theta(\os P \cs)
\end{eqnarray}

\noindent
We will say that an encoding that satisfies (\ref{parallel pres}) 
and (\ref{renaming pres})
is {\it uniform}\footnote{The definition of uniformity has 
emerged from discussion with Iain Phillips and Maria Grazia Vigliotti.
They pointed out the necessity of the condition  
$\forall i\in N\; \sigma(i) = \theta(i)$ in order 
for the encoding to preserve symmetry, and hence for Corollary~\ref{noenc1} to hold.
They also pointed out the necessity of the condition about connectivity in 
Corollary~\ref{noenc2}.}.

Concerning the notion of semantics, we call ``reasonable'' 
a semantics which distinguishes two processes $P$ and $Q$ whenever 
there exists a maximal (finite or infinite) 
computation of $P$ in which the intended observables (some visible 
actions) are different from the observables in any (maximal) 
computation of $Q$. 
In the following, our intended observables are the 
actions performed on channel {\it out}.
Note that the above condition cannot be 
satisfied by a semantics which is insensitive to 
infinite $\tau$ loops, such as weak bisimulation 
or coupled bisimulation. 

\begin{corollary}\label{noenc1}
There exist no uniform encoding of $\pi_m$
into  $\pi_s$ preserving a  reasonable semantics.
\end{corollary}

\noindent
{\bf Proof}
Uniformity preserves symmetry, and a reasonable semantics 
distinguishes an electoral system from a non-electoral one.
Hence apply Theorem~\ref{nonexistence api} and Proposition~\ref{quadrato}.
Note that the hypergraph in Proposition~\ref{quadrato} has 
one-orbit automorphisms.
\qed

For the next result we need a further condition: We say that an encoding 
$\os\cdot\cs$ {\it does not increase the level of connectivity of a network} 
if for all processes $P$ and $Q$, if ${\it fn}(P) \cap {\it fn}(Q) = \emptyset$, 
then ${\it fn}(\os P \cs) \cap {\it fn}(\os Q \cs) = \emptyset$.

\begin{corollary}\label{noenc2}
There exist no uniform encoding of $\pi_m$
into CCS$_{vp}$ or into $\pi_I$ which does not increase the level of connectivity and 
which preserves a reasonable semantics.
\end{corollary}

\noindent
{\bf Proof}
Analogous, by Theorem~\ref{nonexistence CCS} and Proposition~\ref{quadrato}.
Note that the hypergraph in Proposition~\ref{quadrato} has 
well-balanced automorphisms satisfying the conditions of 
Theorem~\ref{nonexistence CCS}.
\qed

Note that if we relax condition (\ref{parallel pres}), imposing just 
generic compositionality  instead, i.e. 
\begin{eqnarray}\label{comp}
\os P\,|\,Q\cs &=&C[\;\os P\cs, \os Q\cs\;]
\end{eqnarray}

\noindent 
with $C[\cdot,\cdot]$ generic context, 
then these non-encodability results do 
not hold anymore. In fact, we could 
give an encoding of the form 
\begin{eqnarray*}
\os P\,|\,Q\cs &=&\nu y_1\nu y_2\ldots \nu y_n  (\os P\cs \,|\, M \,|\, \os Q\cs)
\end{eqnarray*}

\noindent 
where $M$ is a ``monitor'' process  which coordinates the activities of 
$P$ and $Q$, interacting with them via the fresh channels 
$y_1,y_2,\ldots,y_n$. 
The translation of a network $P_1\,|\,P_2\,|\ldots|\,P_n$ would then be 
a tree with the $P_i$'s as leaves, and the monitors as the other nodes. 
The disadvantage of this solution is that it is not a distributed 
implementation; on the contrary, it is a very centralized one. 

\section{Discussion and future work}

The non-existence results of this work 
hold even if we disregard {\it unfair} computations.
The proof of Theorem~\ref{nonexistence api} in fact 
can be slightly 
modified so that for the construction of $C_{h+1}$ from $C_h$ 
we consider each time a different process in the network. 
In this way, the limit of the sequence is a fair computation.

Our Theorems~\ref{nonexistence api} and \ref{nonexistence CCS}
correspond to Theorems 3.2.1 and 4.2.1 in \cite{Bouge:1988:AI},
for $CSP_{\it in}$ and $CSP$ respectively.
The main difference with those results
is that here we are dealing with much richer  
languages. In particular, both the 
$\pi$-calculi and CCS$_{vp}$ admit the parallel operator inside
every process, and not just at the top-level as it is the case 
for $CSP_{\it in}$ and $CSP$ (at least, for the versions considered in 
\cite{Bouge:1988:AI}: all processes in a network are strictly sequential). 
This leads to an essential difference. Namely,  
the proof of Boug\'e shows that the network can get stuck in the attempt 
to elect a leader: since an output action in $CSP_{\it in}$
can be only sequential, the prefix of a computation which leads to the 
first output action, repeated by all processes, 
brings to a global deadlock.
Our proof, on the contrary, 
shows that the system can run forever 
without reaching an agreement: 
whenever a first output action occurs, 
all the other processes can
execute their corresponding output action as well, 
and so on, thus generating an infinite computation 
which never breaks the symmetry.
Another difference is that in the $\pi$-calculus the 
network can evolve dynamically. This is the reason 
why Theorem 4.2.1 in \cite{Bouge:1988:AI} does not hold for 
$\pi_m$ (as shown by our Theorem~\ref{existence pi}).
This feature complicates the proof of 
Theorems~\ref{nonexistence api} since we have to take 
into account the evolution of the 
automorphism. 

The use of the parallel operator as a free 
constructor usually enhances significantly 
the expressive power of 
a language. It is for instance essential for implementing  
choice (at least in a restricted form). 
In fact, Boug\'e (1988) has shown 
that it is not possible to encode 
$CSP_{\it in}$ into $CSP_{no}$
(the sublanguage of $CSP$ with neither input nor output guards
in the choice), 
while Nestmann and Pierce (2000) have shown that 
the $\pi$-calculus with input-guarded choice ($\pi_i$)
can be embedded into the $\pi$-calculus without choice 
($\pi_a$). The crucial point is that 
the parallel operator 
allows to represent the main characteristic of the
choice, namely the simultaneous availability of 
its guards.

Sangiorgi and Walker (2001) cite our separation result between 
$\pi_m$ and $\pi_s$
wrt a slightly different semantic condition, 
which is the following: 
\vspace{.1in}
\begin{center}
\begin{minipage}{4.5in}
{\em
  For any $P$ and any $N\subseteq{\it fn}(P)$, if every maximal 
  computation of $P$ contains exactly one action whose subject is in $N$, 
  then every maximal computation of $\os P\cs$ contains exactly one action whose 
  subject is in $N$.
}
\end{minipage}
\end{center}
\vspace{.1in}
Strictly speaking, with this condition 
the separation result does not 
follow from Theorem~\ref{nonexistence api}. The problem is 
that our notion of electoral systems requires 
all processes to execute the action $\overline{out}\,n$ 
after the leader is elected. 
This requirement corresponds to imposing that all processes 
will eventually know whom the winner is, which is a standard 
condition in the notion of electoral system found in literature. 
However, we could have considered
a simpler (more permissive) notion of electoral system, 
obtained by requiring, in Definition \ref{electoral}, 
that $C'$ contains only one action of the form 
$\overline{\it out}\ n$ (performed, presumably, by the winner). 
All results presented in this paper would remain valid 
under this new notion of electoral system, and in this way 
the separation result could be proved also wrt the 
above variant of the semantic condition.

One way to interpret the results presented in this paper
is that mixed choice is a
really difficult mechanism to implement.
The only possibility to achieve a fully distributed and symmetry-preserving 
implementation probably is to use randomized techniques. 
Francez and Rodeh (1980) \nocite{Francez:80:FOCS}
have proposed a randomized implementation of CSP, 
however, their solution it is not fully
satisfactory because it is not robust wrt adverse scheduling strategies. 
We are currently investigating a probabilistic extension of $\pi_i$
for this purpose, called $\pi_{pa}$ \cite{Herescu:00:FOSSACS}. 
We have shown that in $\pi_{pa}$ it is possible to express the solution 
to some of the leader election problems that would not be solvable in 
$\pi_i$ (or $\pi_s$), so this seems encouraging. 

\subsection*{Acknowledgements}
I would like to thank the following people 
for stimulating 
and insightful discussions: 
Ilaria Castellani, Pat Lincoln,
Dale Miller, Uwe Nestmann, 
Prakash Panangaden,
Benjamin Pierce, 
Rosario Pugliese, Julian Rathke, Davide Sangiorgi, 
Scott Smolka and Eugene Stark. 
Special thanks to Wu Peng for pointing out errors 
(now corrected)
in Definitions~(\ref{algorithm}) and (\ref{alg_II}).
I gratefully acknowledge also Iain Phillips and Maria Grazia Vigliotti 
for pointing out that a previous notion of uniform embedding was problematic, 
and for helping to develop the present formulation. 
They also pointed out the necessity of the condition about connectivity in 
Corollary~\ref{noenc2}.
Furthermore I would like to express 
my gratitude to the anonymous reviewers, whose careful 
and thorough reports have helped enormously to improve the paper.
Last but not least, I would like to thank the editors 
of this special issue of {\it MSCS}, Luca Aceto, Giuseppe Longo
and Bj\"orn Victor, for the excellent job they did as editors.

\bibliographystyle{apalike}
\bibliography{../biblio_cat}

\end{document}